\documentclass[traditabstract]{aa}

\usepackage{graphicx}
\usepackage{subfig}
\usepackage{float}
\usepackage{times}
\usepackage{natbib}
\usepackage{todonotes}

\newcommand{\enzo}{{ENZO\ }}

\newcommand{\athena}{{\it Athena}}

\begin{document}
 
\title{Detecting shocked intergalactic gas with X-ray and radio observations}

\author{F. Vazza\inst{1,2,3}, S. Ettori\inst{4,5}, M. Roncarelli\inst{1,4}, 
M. Angelinelli\inst{1},  M. Br\"{u}ggen\inst{2}, C. Gheller\inst{6}}

\offprints{%
 E-mail: franco.vazza2@unibo.it}
\institute{Dipartimento di Fisica e Astronomia, Universit\'{a} di Bologna, Via Gobetti 93/2, 40122, Bologna, Italy
\and  Hamburger Sternwarte, Gojenbergsweg 112, 21029 Hamburg, Germany
\and Istituto di Radioastronomia, INAF, Via Gobetti 101, 40122, Bologna, Italy
\and INAF, Osservatorio di Astrofisica e Scienza dello Spazio, via Pietro Gobetti 93/3, 40129 Bologna, Italy
\and INFN, Sezione di Bologna, viale Berti Pichat 6/2, I-40127 Bologna, Italy
\and Swiss Plasma Center, EPFL, SB SPC Station 13 - 1015 Lausanne, Switzerland }

\authorrunning{F. Vazza, S. Ettori, M. Roncarelli, et al.}
\titlerunning{Shocked gas in radio and X-ray observations}

\date{Accepted ???. Received ???; in original form ???}

\abstract{Detecting the thermal and non-thermal emission from the shocked cosmic gas surrounding large-scale structures represents a challenge for observations, as well as a unique window into the physics of the warm-hot intergalactic medium. 
In this work, we present synthetic radio and X-ray surveys of large cosmological simulations in order to assess the chances of jointly detecting the cosmic web in both frequency ranges. We then propose best observing strategies tailored for existing (LOFAR, MWA and XMM) or 
future instruments (SKA-LOW and SKA-MID, \athena\ and eROSITA). We find that the most promising targets are the extreme peripheries of galaxy clusters in an early merging stage, where the merger causes the fast compression of warm-hot gas onto the virial region.  By taking advantage of a detection in the radio band, future deep X-ray observations will probe this gas {\it in emission}, and help us to study plasma conditions in the dynamic warm-hot intergalactic medium with unprecedented detail.}
\maketitle

\label{firstpage}
\begin{keywords}
galaxy clusters, ICM
\end{keywords}

\section{Introduction}
\label{sec:intro}
Numerical simulations  \citep[e.g.][]{1999ApJ...514....1C,2001ApJ...552..473D,2016MNRAS.462..448G,2018arXiv181001883M} have shown that the most important mass component of the  baryons in the cosmic web is the elusive Warm-Hot Intergalatic Medium (WHIM), a rarefied gas with densities $\rm n \sim 10^{-5}-10^{-4} \rm ~part/cm^3$ and temperatures $\rm T \sim 10^{5}-10^{7} \rm K$. 

The WHIM should fill the volume within cosmic filaments as well as in the outskirts of galaxy clusters, and attain its temperature via  strong ($\mathcal{M} \gg 10$) accretion shocks.

Detections of absorption lines through the WHIM of intracluster filaments have been claimed \citep[e.g.][]{2008A&A...482L..29W,2010ApJ...715..854N}.  More recently, \citet{2018Natur.558..406N} reported the possible detection of two OVII absorbers  in the X-ray spectrum of a quasar at $z \geq 0.4$, possibly tracing the WHIM.\\

In a few nearby galaxy clusters and limited to  few narrow sectors, the thermodynamical properties of the intracluster medium have been mapped with X-ray observations out to $R_{\rm 100}$  \citep[e.g.][]{si11,ur11}. Others mapped the gas properties in concentric rings out to $R_{\rm 200}$ by combining X-ray and SZ data \citep[][]{ghi19,ett19,eck19}. The ends of 
five massive filaments connected to the massive cluster A2744 have been observed with XMM-Newton  \citep[][]{2015Natur.528..105E}, possibly representing the first images of cosmic filaments in the X-ray band. This made it possible to estimate that $5-10\%$ of the mass fraction of missing baryons may be bound in such objects. 

Moreover, the study of the 
 Sunyaev-Zeldovich effect from the outer region of clusters, either in single pointings of interacting clusters \citep{2013A&A...550A.134P,2018A&A...609A..49B} 
 or in stacked observations of larger samples \citep{2017arXiv170910378D,2018arXiv180504555T} has detected the hot ( $\sim 10^{7} \rm ~K$) and very overdense ($\sim 10-10^2 \rho/\langle \rho \rangle$) gas component, potentially 
 contributing to $\sim 10-50 \%$ of missing cosmic baryons.\\

Cosmological hydrodynamical simulations have shown that the WHIM follows thde underlying galaxy distribution \citep[see, e.g.,][]{nevalainen15} and that its diffuse emission is responsible for a significant fraction of the unresolved X-ray background in very deep Chandra observations \citep[see][]{roncarelli06a,hickox07}, with a predicted surface brightness of the order of $1-5 \cdot  10^{−13} \rm erg/(cm^2 ~s ~deg^2)$ in the $\sim 0.5-1 ~\rm keV$ energy band, with uncertainties related to its metal composition \citep[][]{ursino10,2011ApJ...731...11C,2012MNRAS.424.1012R}.
However, the systematic detection and characterisation of single WHIM systems remain a challenge due to its low emissivity. Hence, efforts have focused on stacking and statistical studies on the (auto)correlation function of its X-ray signal \citep[see, e.g.,][]{piro09,takei11,ursino11,cappelluti12,kolodzig18}. \\

The planned X-ray mission \athena\ X-ray observatory  \footnote{http://www.the-athena-x-ray-observatory.eu}, expected to be launched by $\sim 2030$, holds great promise to detect the WHIM in absorption.  
Among its ambitious goals, \athena\ aims to trace the missing baryons in the intergalactic medium via detecting their absorption lines,  through the emission of bright sources up to $z\sim 2$.   It is expected that in total $\sim 80 $ sources can be studied at the highest possible resolution for spectroscopic studies with the instrument X-IFO ($\Delta E=2.5 ~ \rm eV$) \citep[][]{2012arXiv1207.2745B}.
 The limitation of this technique is that, of course, most of these sources are unpredictable and variable, and only a dozen of bright enough sources per year may be detected to study filaments.  
 Concepts for future high-resolution X-ray imagers, capable of further reducing the limiting effect of the X-ray background from unresolved point-like sources have recently been presented in the context of the US Decadal Survey (e.g. Lynx \citealt{2018arXiv180909642T}, AXIS \citealt{2019arXiv190304083M}).
 
In this work, we wish to explore the complementary approach, in which {\athena} detects the WHIM in emission. 

\citet{2018MNRAS.476.4629P} have recently presented detailed radiative transfer calculations of the X-ray emission from a cluster at $z \approx 0.3$, reporting that the emission from the WHIM only accounts for the $\sim 5\%$ in the [0.5-2] keV band and $\sim 1\%$  in the [2-10] keV band. In general, the WHIM is found to have a more filamentary structure than the $\geq 10^7 ~\rm K$ gas phase, extending several $\rm Mpc$ out from the virial regions of galaxy clusters. 
Using a larger volume with the Illustris-TNG suite, \citet{2018arXiv181001883M} recently studied the WHIM properties from $z=4$ to $z=0$, confirming that most of 
that filaments are more baryon-rich than the cosmic average, but that they have a significantly lower metallicity than the ICM, which makes their observability via X-ray observations challenging.\\

In addition to the X-ray window, also the radio window maybe able to image the cosmic web, thanks to the current (e.g. LOFAR, MWA, ASKAP, MEERKAT) and future (e.g. the Square Kilometer Array) generations of large radio telescopes. 
Cosmological simulations have shown that filaments of the cosmic web are surrounded by strong and quasi-stationary accretion shocks \citep[e.g.][]{ry03,pf06,va11comparison}, at which a tiny fraction of relativistic electrons may be accelerated. This is similar to what occurs in radio relics or cluster radio shocks at the periphery of clusters \citep[e.g.][]{hb07,wi17er,2019arXiv190104496V}. 

A few radio observations have already claimed  the detection of diffuse synchrotron emission from the shocked gas at the interface between galaxy clusters and filaments attached to them  \citep[][]{2002NewA....7..249B,2010A&A...511L...5G,2013ApJ...779..189F,2018MNRAS.tmp.1093V}. Moreover,  the observation of Faraday Rotation by filaments in the Coma cluster has been claimed \citep[][]{bo13}. Recently, the signature in Faraday space of filaments overlapping with the emission of a $z=0.34$ radio galaxy has been claimed by 
\citet[][]{2018arXiv181107934O}. The detection of faint diffuse radio emission at the interface of pre-merger galaxy clusters has been reported using LOFAR-HBA by \citet[][]{2018MNRAS.478..885B} and Govoni et al. (submitted).

Simulations have shown that the low surface brightness ($\leq \rm \mu Jy/arcsec^2$ at $\sim 100 ~\rm MHz$ ), highly polarised ($\sim 70\%$)  and  large angular scale  ($\geq 1^{\circ}$) emission that is expected to be produced by the shocked cosmic web  \citep[e.g.][]{2004ApJ...617..281K,2011JApA...32..577B,va15radio} makes the low-frequency radio spectrum ($\nu \leq 300 \rm ~MHz$) the most suitable for a detection, owing to the typically superior sampling of short baselines in low-frequency radio telescopes. In particular, the radio continuum surveys of  SKA-LOW should detect parts of the magnetic cosmic web, with statistics depending on the (unknown) details of particle acceleration and magnetic field distribution in such rarefied plasma \citep[e.g.][]{va15ska,va15radio,va17cqg}. Additional to this, also polarisation surveys with the SKA-MID may be able to detect the Faraday Rotation signal from the terminal part of filaments connected to massive galaxy clusters, provided that a large statistics of polarised sources is available \citep[][]{lo18}. \\
 
The exciting possibility of detecting both {\it thermal} and {\it non-thermal} emission from the cosmic web with joint X-ray and radio surveys is the subject of this work.
In a pilot study for the "SKA-Athena Synergy White Paper" (\citealt{athena_ska}, Sec.~5.2.1) we first investigated the potential for a synergy between SKA and \athena\ in the study of the rarefied cosmic web. Our first results suggest that cluster outskirts are promising targets, with a small but non-negligible fraction of the cosmic web that might be detectable by both instruments (working at their nominal maximum capabilities). This possibility will make it possible to study the WHIM with \athena\ not only via absorption lines towards high-$z$ powerful sources, but also on a few, carefully selected objects. 
Based on these results, we use one of the largest cosmological magneto-hydrodynamical simulations ever produced to assess to which extent future X-ray and radio observations can constrain the physical properties of the WHIM. 

The structure of the paper is as follows:
in Sec.~\ref{sec:methods} we introduce our simulations, and in particular in Sec.~\ref{subsec:sky} we introduce our methods to produce sky models of our simulated universes. In Sec.~\ref{sec:results} we present our results for the intrinsic emission properties of the cosmic web in X-ray and radio bands, while in Sec.~3.3.3 we specifically investigate mock \athena\ and SKA observations of our fields, and in Sec.~\ref{subsubsec:sixte} we present preliminary simulations of future spectroscopic analyses with X-IFU. Physical and numerical limitations of our results are given in Sec.~\ref{sec:discussion} before we conclude in Sec.~\ref{sec:conclusions}. 

 \begin{figure}
  \includegraphics[width=0.49\textwidth]{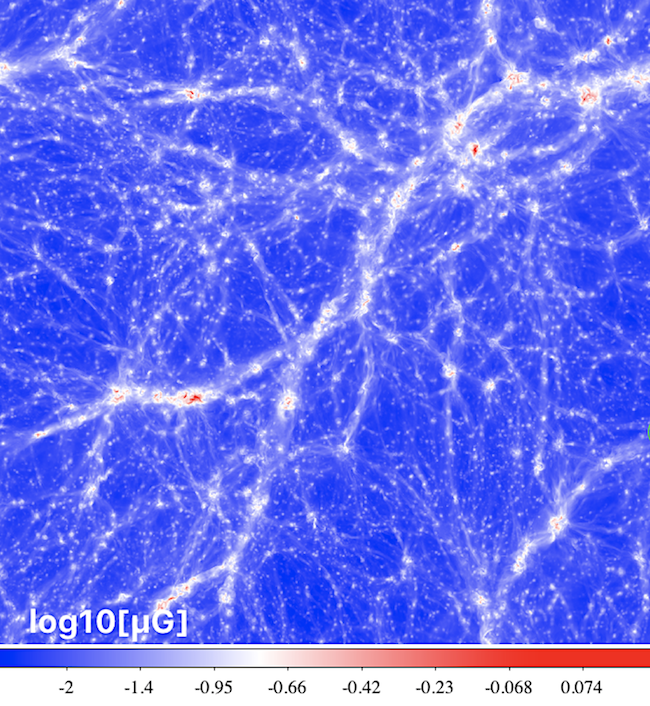}
\caption{Projected (mass weighted) magnetic field strength at $z=0.05$ for our simulated $100^3 ~\rm Mpc^3$ volume.} 
\label{fig:Bproj}
\end{figure}

 \begin{figure*}
  \includegraphics[width=0.99\textwidth]{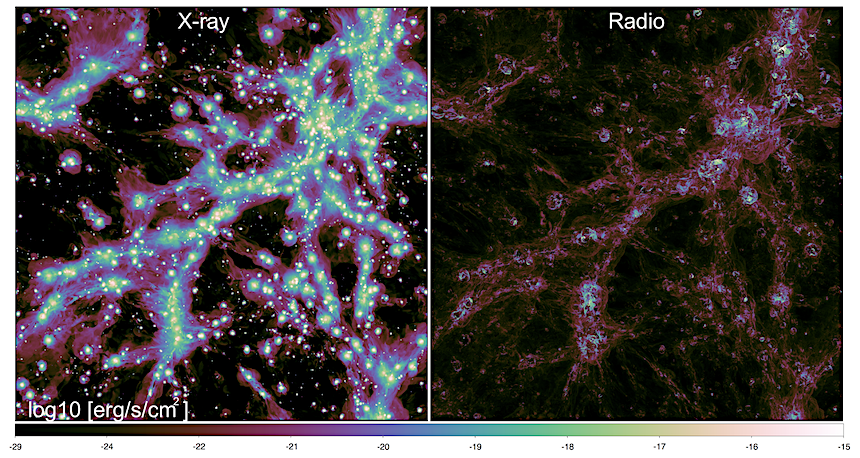}
\caption{Projected X-ray emission in the [0.8-1.2] keV band (left) and  mock radio emission at 260 MHz (right) for our $100^3 ~\rm Mpc^3$ volume located at $z \approx 0.05$.}
\label{fig1}
\end{figure*}

\section{Methods}
\label{sec:methods}
\subsection{Cosmological simulations}
As in previous work \citep[][]{va17cqg,va18frb}, we simulated a comoving  $100^3 \rm Mpc^3$ box with a uniform grid of
$2400^3$ cells and $2400^3$ Dark Matter particles, using the cosmological
MHD code  {\enzo} \footnote{www.enzo-project.org} \citep{enzo14}. 
The fixed (comoving) spatial resolution of this run is $\Delta x=41.6 ~\rm kpc/cell$ 
while the fixed Dark Matter
mass resolution is $m_{\rm dm} = 8.62 \cdot 10^{6} M_{\odot}$ 
per particle. We initialized magnetic fields at $z=45$
as a simple uniform background of 
of $B_0=0.1 \rm nG$ (one order-of-magnitude below the upper limits on primordial magnetic fields from the analysis of the CMB, \citealt[][]{PLANCK2015}), and we used the MHD method of  Dedner \citep[][]{ded02}, ported to GPUs \citep[][]{wang10} to evolve magnetic fields at run-time.

Our  run is non-radiative and does not include any treatment for star formation or feedback from active galactic nuclei. To a first approximation, these processes are not very relevant for the radio and X-ray properties of the peripheral regions of galaxy clusters and filaments, which are our main focus (see Sec.~4 for a discussion). 

We assumed a $\Lambda$CDM cosmological model, with density parameters $\Omega_{\rm BM} = 0.0455$, $\Omega_{\rm DM} =
0.2265$ (BM and DM indicating the baryonic and the dark matter respectively),  $\Omega_{\Lambda} = 0.728$, and a Hubble constant $H_0 = 70.2 \rm ~km/(sec ~ Mpc)$. 
In Sec.\ref{subsubsec:sixte} we will also present results for the resimulation of a massive galaxy clusters with the same setup, but using nested initial conditions and Adaptive Mesh Refinement (8 levels) to achieve a higher resolution ($\Delta x_8 \approx 4 \rm ~kpc/cell$), similar to  \citet{va18mhd}. 

An impression of the three-dimensional distribution of magnetic fields in our simulation (which is one of the biggest MHD simulations ever performed in cosmology) is given in Fig.~\ref{fig:Bproj}, and shows the variety of magnetic field strengths and configurations at $z=0.05$. 

\subsection{Sky models}
\label{subsec:sky}
\subsubsection{X-ray emission}
\label{subsubsec:xray}

For the X-ray emission, we assumed for simplicity a single temperature and a single (constant) composition for every cell in the simulation, and we  compute the emissivity from the B-APEC emission model {\footnote{https://heasarc.gsfc.nasa.gov/xanadu/xspec/manual/Models.html}}, assuming ionization equilibrium and including continuum and line emission.

We consider a {\it constant} metallicity across the volume, $Z/Z_{\odot}=0.3$ (see a discussion in Sec.~4 for the rather small impact of metallicity in most of our estimates). 

For each energy band, we compute the cell's X-ray emissivity, $S_X=n_H n_e \Lambda(T,Z) dV$, where $n_H$ and $n_e$ are the number density of hydrogen and electrons (assuming a primordial composition) respectively, and $dV$ is the constant volume of our cells.

We do not include the additional contribution from the Inverse Compton emission from the same relativistic electrons accelerated by shocks and responsible for the radio emission (see next Section), whose amplitude depends on the assumed electron energy distribution at low-energies \citep[e.g.][]{2015A&A...582A..20B}. However, our estimates show that the Inverse Compton in the [0.8-1.2] keV band is negligible ($\leq 1\%$) compared to the thermal emission of the cluster. It may start to dominate only at very large radii, $\geq 2-3 R_{\rm 100}$, at which no detection seems to be feasible with realistic exposure times.

The left panel of Fig.~\ref{fig1} shows the integrated X-ray emission from the simulated box located at at $z=0.05$ , in the [0.8-1.2] keV energy band. 
At the angular distance corresponding to this redshift ($D_A \approx 201.6 \rm Mpc$) this volume covers $28.4^{\circ} \times 28.4^{\circ}$.
For reference, \athena's Wide Field Imager field of view is $\sim 40' \times 40'$ , while X-IFU's field of view is $\sim 5' \times 5'$.

\subsubsection{Radio}

\label{subsubsec:radio}

We predict the synchrotron radio emission assuming that diffusive shock acceleration (DSA, e.g. \citealt{ka12} and references therein) is able to accelerate a very small fraction of thermal electrons swept by shocks up to relativistic energies ($\gamma \geq 10^3-10^4$), and that the intergalactic medium has a non-negligible magnetic field, as suggested by our MHD simulation (e.g. Fig.\ref{fig:Bproj}).
As in previous work \citep[][]{va15radio}, we identify shocks in the simulation in post-processing with a velocity-based approach, and we compute the radio emission from electrons accelerated in the shock downstream following \citet{hb07}.  The typical efficiency ($\xi_e$) considered in the conversion efficiency from shock kinetic energy into the energy of relativistic electrons is small and it scales with the Mach number and the upstream gas temperature as in \citet{hb07}: for example it is $\xi_e \approx 10^{-6}$ for $\mathcal{M}=3$ shocks in a $T=10^7 \rm K$, and $\xi_e \approx 6 \cdot 10^{-4}$ for $\mathcal{M} \ge 50$ shocks with a $T=10^{5} \rm K$.
The additional (possible) role of shock obliquity \citep[e.g.][]{wi17er} and of fossil reaccelerated electrons \citep[e.g.][]{2013MNRAS.435.1061P} is neglected here for simplicity; we caution however that the additional presence of fossil electrons in cluster outskirts and in filaments will increase our estimates here, at least limited to $\mathcal{M} \leq 3-4$ shocks in the simulation (while for stronger shocks the direct injection from DSA should dominate the emission in any case). 
The downstream radio emission is the convolution of the several power-law distributions of electrons which overlap in the cooling region,  to which we assign the integrated radio spectrum of $I(\nu) \propto \nu^{-s}$, where $s=(p-1)/2+1/2$, with $p=2(\mathcal{M}^2+1)/(\mathcal{M}^2-1)$. \\

An example of the radio emission at $\nu =260 ~\rm MHz$ from our simulation is given in Fig.~\ref{fig1}.
The SKA-LOW primary beam at this frequency should be of order $\sim 5^{\circ} \times 5^{\circ}$, while our sky model covers $28.4^{\circ} \times 28.4^{\circ}$. The radio emission is clearly more diffuse compared to the X-ray emission, because unlike the latter it does not scale (only) with gas density, but with the shock kinetic power, which can be significant in cluster outskirts \citep[e.g.][]{ry03}. The fact that the radio power typically extends out to larger cluster radii makes it a very good probe of the rarefied cosmic web, but at the same time reduces the chances of overlap with X-ray detections, as we shall see in the next Section,  with a sweet spot on the scale of cluster outskirts.

 \begin{figure}
  \includegraphics[width=0.45\textwidth]{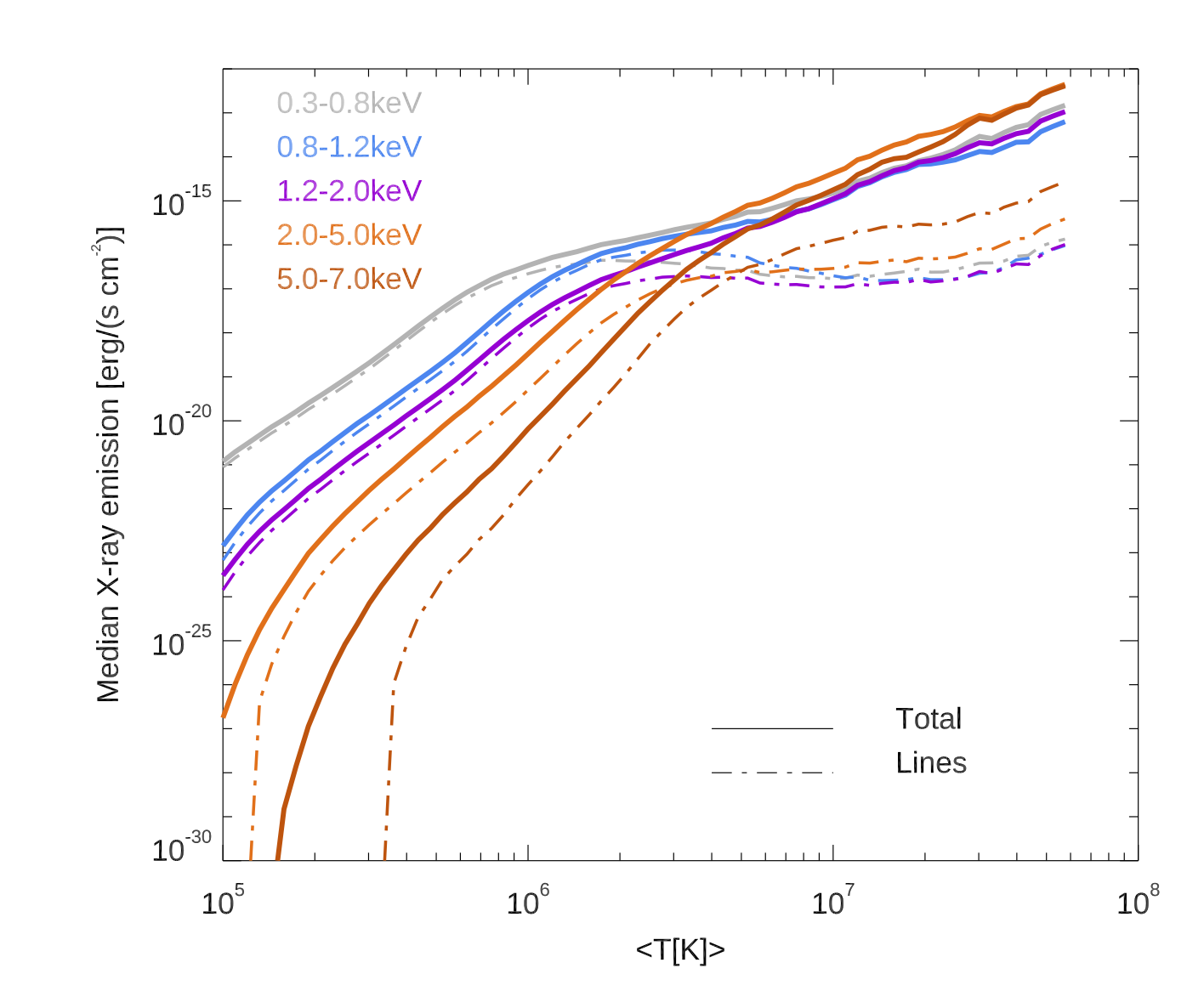}
  \includegraphics[width=0.45\textwidth]{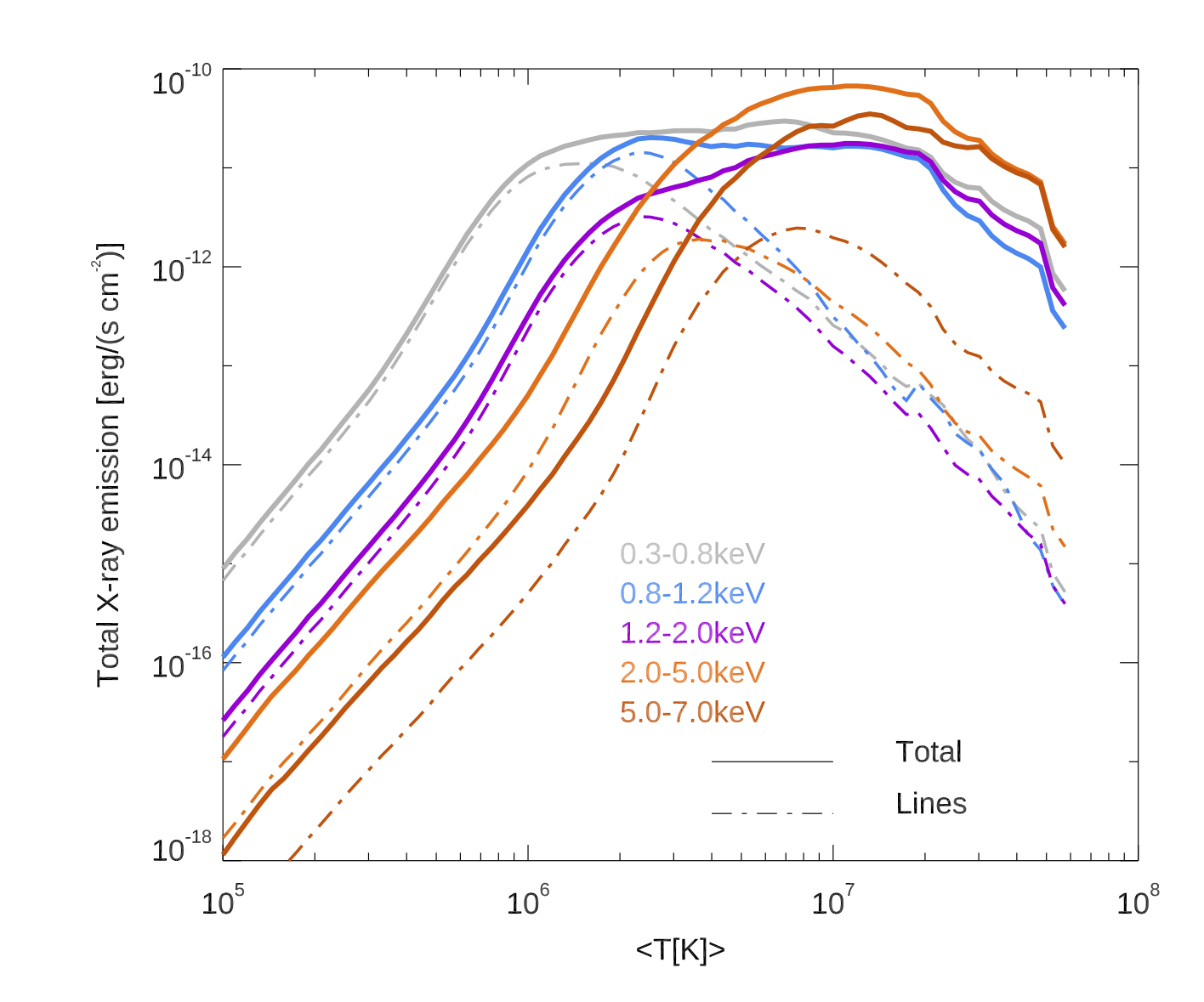}
   \caption{Distribution function of the median X-ray emission (top) and of the total X-ray emission (bottom) for different energy ranges, as a function of the projected mass-weighted temperature of gas in our simulation. The solid lines give the total X-ray emission, while the dot-dashed lines give the contribution only by line emission (assuming a fixed $0.3~Z_{\odot}$ metallicity everywhere).}
\label{fig:X-ray_bands1}
\end{figure}

 \begin{figure}
  \includegraphics[width=0.495\textwidth]{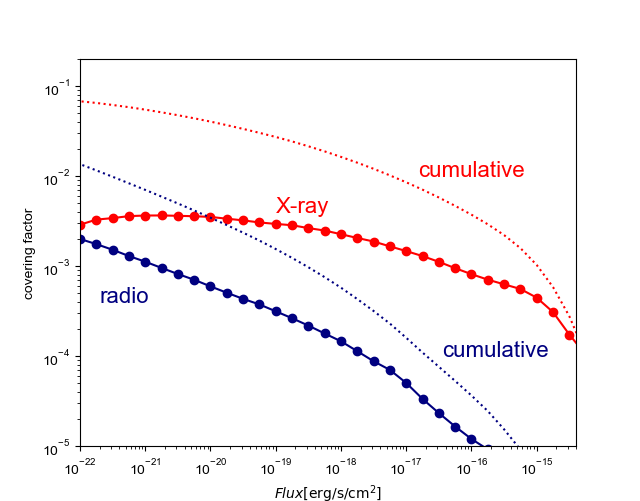}
\caption{Area coverage as a function of X-ray and radio flux (in the [0.8-1.2] keV range and at 260 MHz, respectively) for the same sky models of Fig.\ref{fig:phase}. The solid lines give the differential distribution while the dashed line give the cumulative.}
\label{fig:pdf}
\end{figure}

 \begin{figure}
  \includegraphics[width=0.495\textwidth]{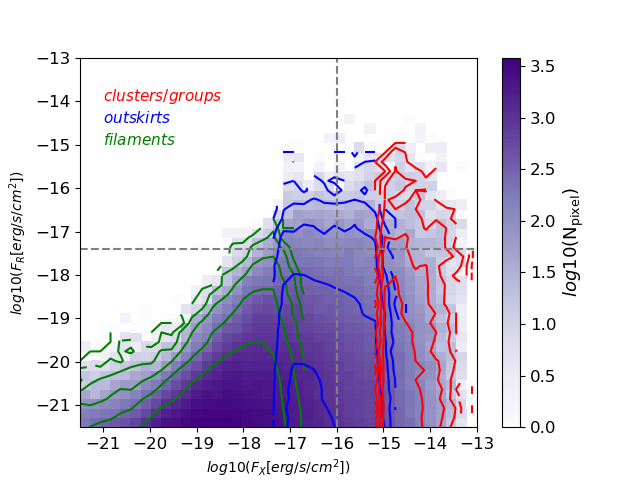}
\caption{Distribution of X-ray and radio flux for all simulated pixels in our sky model (Fig.2).  The contours are equally spaced in logarithmic space ($\Delta log_{\rm 10}=0.5$) and  show the distributions of pixels associated with filaments (green), cluster outskirts (blue) or the virial region of galaxy clusters/groups (red). The additional grey lines delimit the regions which will become observable with future X-ray and radio telescopes.}
\label{fig:phase}
\end{figure}

\section{Results}
\label{sec:results}
\subsection{X-ray emission and radio emission from the cosmic web}

We start by computing the X-ray emission from the entire simulation, as a function of environment and for different energy ranges, namely [0.3-0.8], [0.8-1.2], [1.2-2.0], [2.0-5.0] an [5.0-7.0] keV, assuming $Z=0.3 Z_{\odot}$ everywhere. 
In Fig.~\ref{fig:X-ray_bands1} we show the median and  total X-ray emission from all pixels in the sky model of Fig.\ref{fig1}, binned as a function 
of their gas temperature, which is mass-weighted along the entire line-of-sight {\footnote {It is important to notice here that the {\it average} temperature values along a line of sight of $100$ Mpc
underestimate by a factor $\sim 10$ (or more) the real temperature values in 3D.}}. 

At $T_{\rm mw} \leq 5 \cdot 10^{6} ~\rm K$ (where $T_{\rm mw}$ is the mass-weighted gas temperature), the X-ray emission from the WHIM is always highest in the softest band (0.3-0.8 keV). Most of the emission is free-free radiation above $T_{\rm mw}\geq 10^6 ~\rm K$, while at lower temperatures  emission lines get dominant.  In the entire [0.3-1.2] keV part of the spectrum at these temperatures lines are dominant, and they are $\sim 10$ times more than in the other higher bands (see Sec.\ref{sec:discussion} for variations in the assumed metallicity of our gas). 
Conversely, for $T_{\rm mw} \geq 10^{7} \rm K$, the X-ray emission is always more prominent in the [2.0-5.0] keV band, all contributed by free-free emission. \\

This suggests that, in principle, the [0.3-0.8] keV band has the highest chances of detecting the WHIM in emission. However, when the full response of {\athena} is taken into account, as well as the realistic contribution from astrophysical and instrumental backgrounds, the [0.8-1.2] keV band gives a significantly higher chances of detection. In Sec.\ref{mock} we will show in detail how our modelling suggests the above finding, which stems from a quantitative comparison of the fraction of pixels tracing various environments across our simulation, which can be detected as a function of exposure time and energy bands for different instruments. While we defer to that Section for the demonstration of this finding, we anticipate this result here in order to present more general trends found in our data, limiting for
simplicity to the [0.8-1.2] keV band in particular. \\

Second, we characterized the distribution of X-ray and radio emission from the simulated cosmic web, by comparing the X-ray signal in the [0.8-1.2] keV range to the radio emission at $260$ MHz.
In Figure \ref{fig:pdf} we show the cumulative and differential distribution of X-ray, $F_X$ and radio flux, $F_R$, for the pixels in Fig. 2, in which no observational cuts are taken into account for now. 
While both distribution shows a similar decreased power-law behaviour, the radio distribution has a steeper behaviour for increasing $F_R$, indicating the relative scarcity of powerful detectable radio emission from cluster shocks (e.g. radio relics).
Considering that $\approx 10^{-17}~\rm  erg/s/cm^2$ is the fiducial value for the future sensitivity of SKA-LOW at this frequency, from this plot we can derive that at most a $\sim 1\%$ of the entire radio emission produced by the cosmic web will be detectable even in the future. Due to the flatter slope of the $F_X$ distribution, for a fiducial detection threshold of $\approx 10^{-16}\rm  ~erg/s/cm^2$ we have instead that a $\sim 10\%$ of the total thermal emission from the cosmic web in this energy range may be detectable (assuming observations at high galactic latitudes and long exposures). We will comment more on this issue in Sec.~5.

By computing the distribution of $F_R$ as a function of $F_X$ for the pixels in our simulated sky model (Fig.~\ref{fig:phase}),
 we can demonstrate the lack of a simple relation between the two emission mechanisms. For a given bin in X-ray flux the radio emission can vary by more than 6 orders of magnitude. Interestingly, we measure that the distribution of $F_R$ in the volume peaks for $10^{-17} ~\rm  erg/s/cm^2\leq F_X \leq 10^{-14} ~erg/s/cm^2$ , i.e. of the order the sensitivity level of \athena. 
 
 The pixels in this range of $F_X$ are mostly related to the external accretion regions of galaxy clusters and groups, as shown by the different color-coding adopted in Fig.~\ref{fig:phase}, in which we marked pixels as belonging to clusters or groups in red if they fall within the projected volume ($\leq R_{\rm 100}$) of halos. Those that are at a projected distance $R_{\rm 100} \leq r \leq 3 \cdot R_{\rm 100}$ from halos centres are marked in blue. Finally, pixels were marked in green if they belong to filaments, i.e. if they have a projected gas density in the range $  1 \leq n_{\rm proj}/\langle n \rangle \leq 50 $    and have a projected temperature in the range $10^{5} \rm~ K	\leq T_{\rm mw} \leq 10^7 \rm ~K$.
 
As shown by the additional approximate detection limit for future X-ray and radio observations  ($\sim 10^{-16}\rm  ~erg/s/cm^2$ for [0.8-1.2] keV  and  $4 \cdot 10^{-18} \rm ~erg/s/cm^2$ for $260$ MHz , dashed lines in Fig.\ref{fig:phase}), for the majority of our simulated pixels $F_X \gg F_R$, i.e. the X-ray fluxes are larger than the radio fluxes. The few notable exceptions are the spikes in the radio flux, marked by the vertical stripes in the plot. They are mostly related to temperatures of $T_{\rm mw}\sim 10^6-10^7 ~\rm K$. The visual inspection shows that such patterns are associated with structure formation shocks, often in the periphery of galaxy clusters or in between merging cluster pairs, and we will focus on them in the next section. 

 \begin{figure}
  \includegraphics[width=0.45\textwidth]{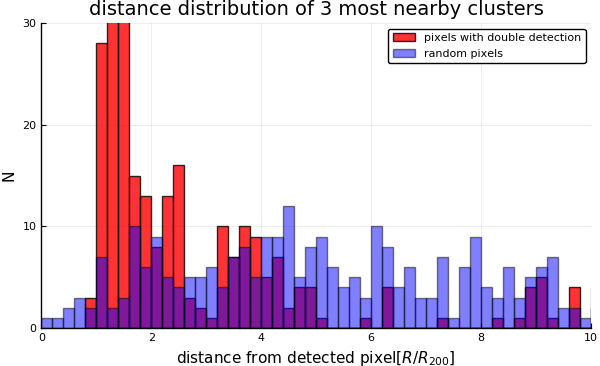}
\caption{ Distribution of the projected distance from doubly detectable pixels of clusters in our simulation, considering the three most nearby clusters from every pixel. The blue histogram gives the distribution of three most nearby clusters around randomly chosen location in the simulation.}
\label{fig:distance_pdf}
\end{figure}

 \begin{figure*}
  \includegraphics[width=0.33\textwidth]{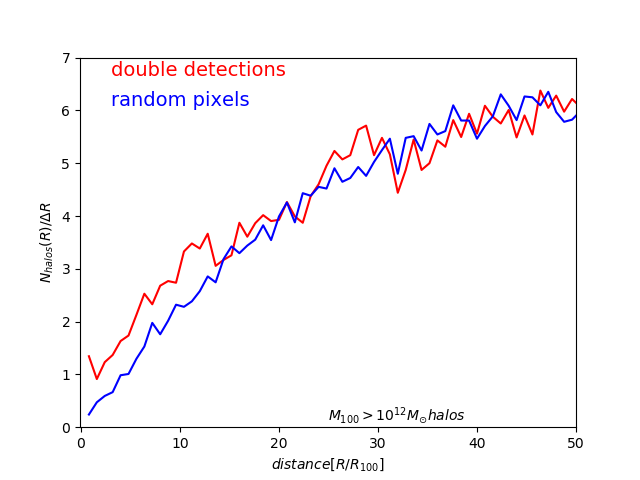}
  \includegraphics[width=0.33\textwidth]{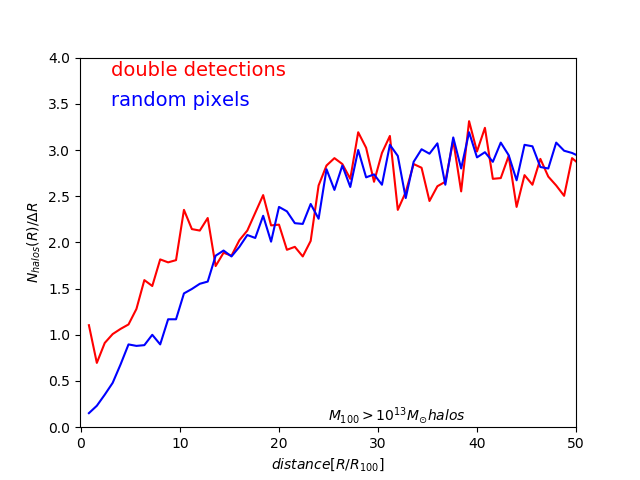}
    \includegraphics[width=0.33\textwidth]{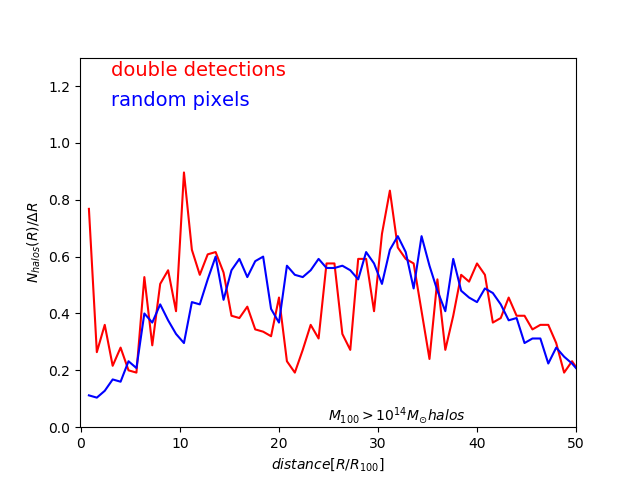}
\caption{Average number of clusters (considering a different lower threshold in mass) at a given distance from the location of joint detections in our simulations.  The blue data are for the distances computed from a randomly drawn set of position, for a number of points equal to our jointly detectable pixels. }
\label{fig:distance_prof}
\end{figure*}

\subsection{Enhanced gas emission from cluster outskirts}

In this section we wish to identify the most promising targets for a joint detection of the WHIM in the radio and in the X-ray band. To this end, we computed the clustering and morphological properties of massive halos found in the neighborhood pixels that are detected in the X-rays and in radio emission.\\

Our $100^3 \rm ~Mpc^3$ simulation box contains a total of 2347 halos with $M_{\rm 100}\geq 10^{12} M_{\odot}$,  268 halos with $M_{\rm 100}\geq 10^{13} M_{\odot}$ and 26 halos with $M_{\rm 100}\geq 10^{14} M_{\odot}$ at $z=0.05$, consistent with the expected mass function \citep[e.g.][]{SH99.1i} within the limits of Poissonian statistics. \\

To quantify the association between clusters in the simulation and pixels detected in X-rays and radio, we start by computing the (projected) distance distribution of the three most nearby clusters around each jointly detectable pixel (see Fig.~\ref{fig:distance_pdf}). We compare this to a similar distribution drawn for an equal number of randomly drawn positions in the simulation. 
On average, the pixels that can be detected, both, in radio and X-rays have a much higher concentration of halos around them. This excess clustering shows up even if we consider the distribution of the ten most nearby clusters around them.
 Likewise, the projected radial distribution of the number of halos (considering different mass cuts) from doubly detected pixels shows a significant excess compared to a randomly drawn distribution, up to $\sim 20-30 ~\rm Mpc$ distances, as shown in Fig.~\ref{fig:distance_prof}.

In order to investigate clusters in the vicinity of these double detectable regions, we computed the  emission of halos at their periphery. 

\begin{figure}
 \includegraphics[width=0.45\textwidth]{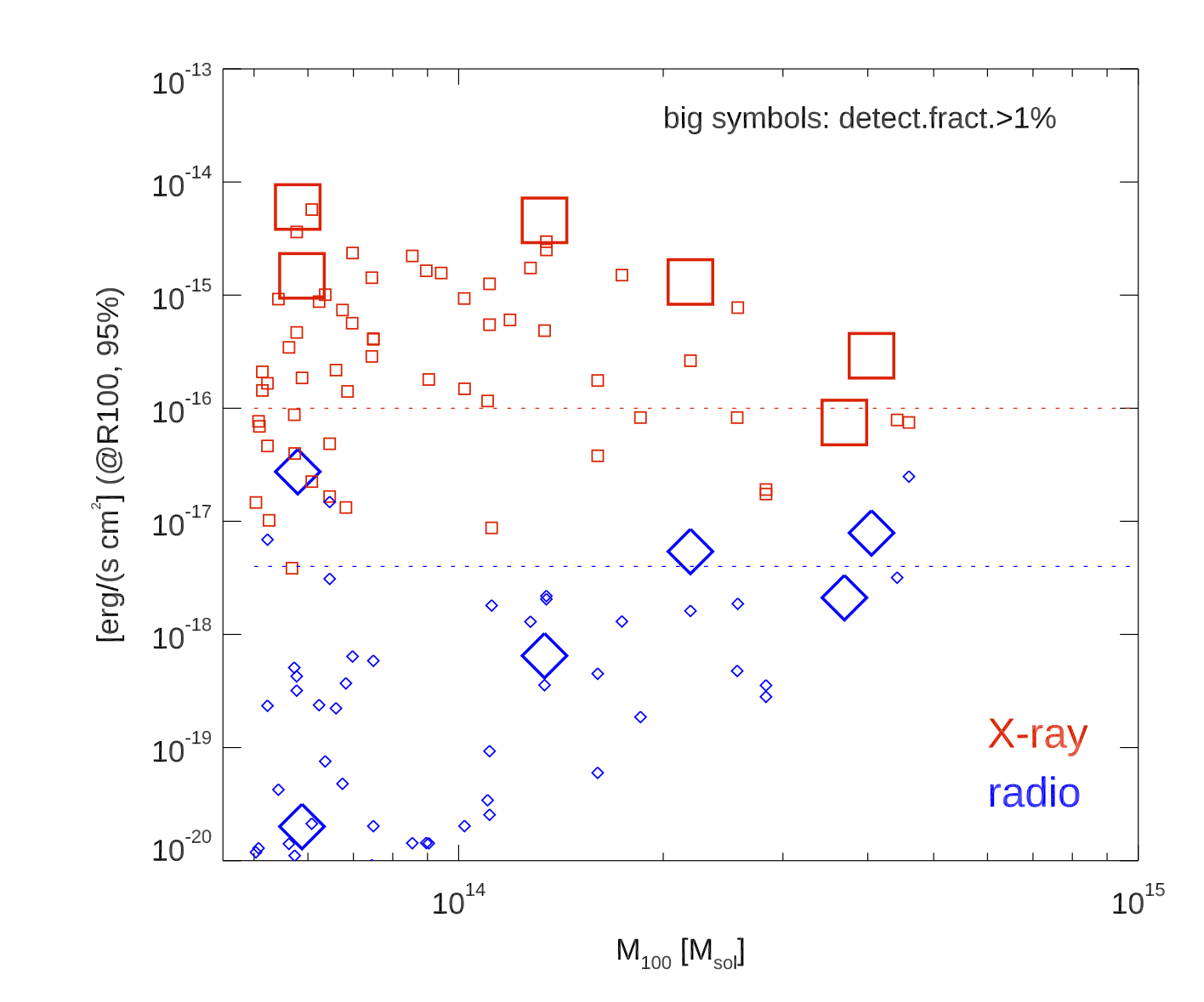}
 \includegraphics[width=0.45\textwidth]{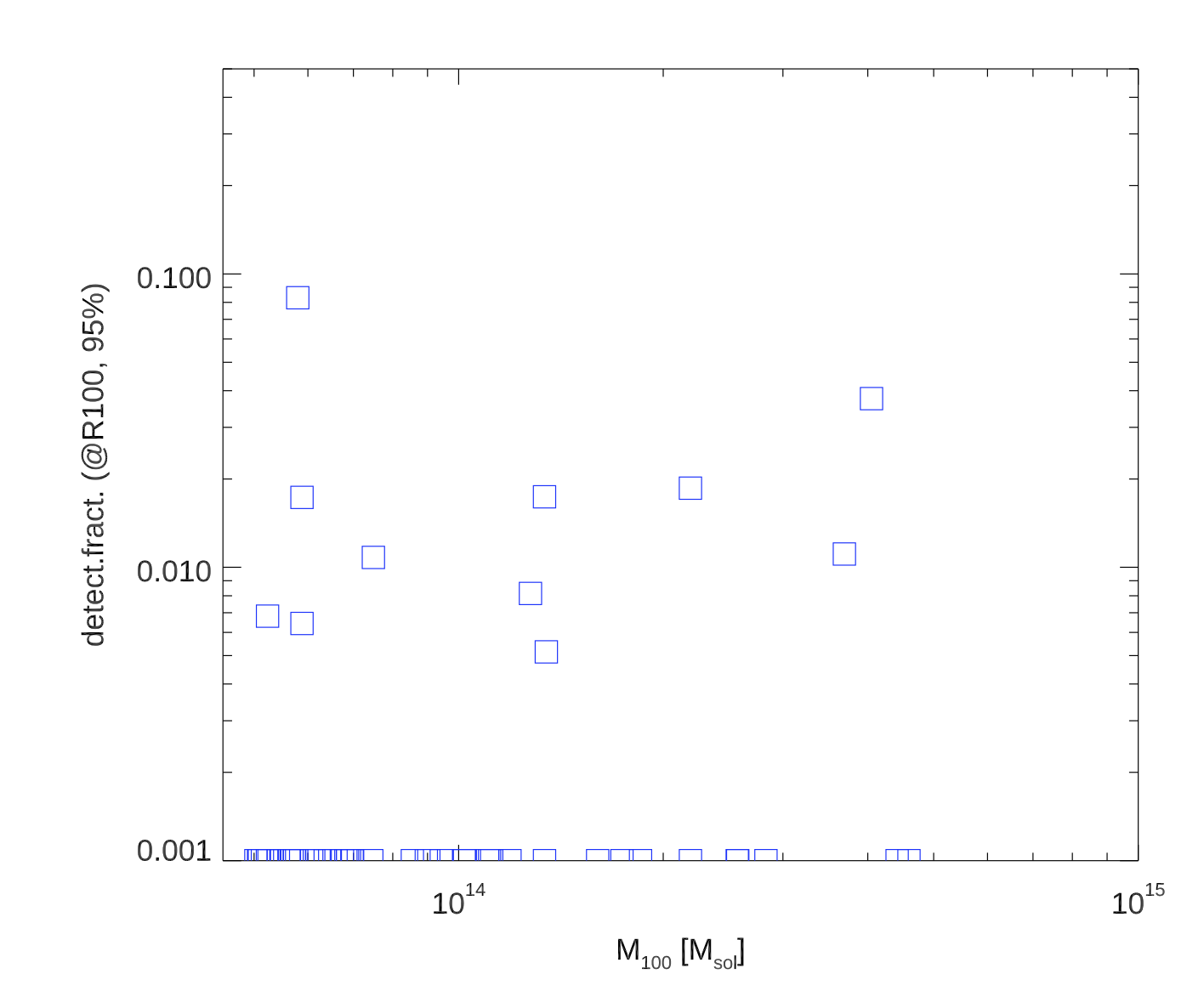}
 \caption{Top panel: 95th percentile of the X-ray and radio emission from a thin radial shell at $R_{\rm 100}$ as a function of the cluster mass, for $\geq 5 \cdot 10^{13} M_{\odot}$ clusters in our simulation at $z=0.05$. The horizontal lines show the reference detection threshold for both cases. Large symbols corresponds to clusters in which $\geq 1 \%$ of the $R_{\rm 100}$ shell can be simultaneously detected in X-ray and in radio. Bottom panel: detection fraction for the $R_{\rm 100}$ shell in our clusters at $z=0.05$, considering the simultaneous detection in X-ray and radio, as a function of the cluster mass}
\label{fig:M100_erg}
\end{figure}

In the top panel of Fig.~\ref{fig:M100_erg} we computed the 95th percentile of X-ray and radio emission at $R_{\rm 100}$ from our set of clusters, and plot it as a function of the cluster total mass within the same radius. Large symbols mark the objects in which at least a $1 \%$ for the area of the $R_{\rm 100}$ shell (considering a width $41.6$ kpc) can be detected simultaneously in X-rays and in the radio band, given our assumed detection thresholds.  At such large distances, the brightest end of the X-ray emission distribution does not scale with cluster mass. The reason is that they are dominated by rare positive brightness enhancements which are mostly due to the presence of large scale companions (whose amplitude does not directly correlate with the host cluster mass), while  the radio emission is overall well correlated with the host cluster mass. 

In our model the radio emission is solely caused by shocks, thus the radio power scales with the  shock kinetic power, $\Phi_K \propto \rho v_s^3 \sim M_{\rm 100}^{3/2}$. Small departures from this relation are due to projection effects, as well as to the non-linear dependence of the acceleration efficiency on the shock Mach number.  In the lower panel of Fig.~\ref{fig:M100_erg} we quantify the detectability of cluster outskirts for each object, by computing the fraction of a shell that can be observed as a function of the host cluster mass: the odds of a joint detection of cluster outskirts do not scale with the cluster mass, and joint detections are found across the entire mass range of our sample.

We checked whether the enhanced emission in cluster outskirts correlate with large-scale disturbances in the matter distribution, using the 
 {\it mass  sparsity}, simply defined as the ratio between the total mass enclosed within two different radii, e.g. $S=M_{\rm 100}/M_{\rm 200}$.  
 While other morphological parameter estimators are useful to characterized the dynamical stage of galaxy clusters in their internal regions, based on X-ray images \citep[e.g.][]{cassano10},
 the mass sparsity correlates well with cosmological parameters ($\sigma_8$ and $\Omega_M$) as well as with the dynamical state of clusters to a larger radius \citep[e.g.][]{2018ApJ...862...40C}, and is a convenient parameter to measure as it involves two integrated quantities (e.g. either the total mass or the integrated X-ray luminosity) which should be robustly probed by future X-ray surveys. 
However, our search for a dependence between the mass sparsity of our cluster sample and enhanced emission in their outskirts did not highlight any dependence as doubly detectable regions can be found at both large ($S \sim 1.6$) and small ($S \sim 1.1$) values of sparsity.\\

\begin{figure}
  \includegraphics[width=0.24\textwidth]{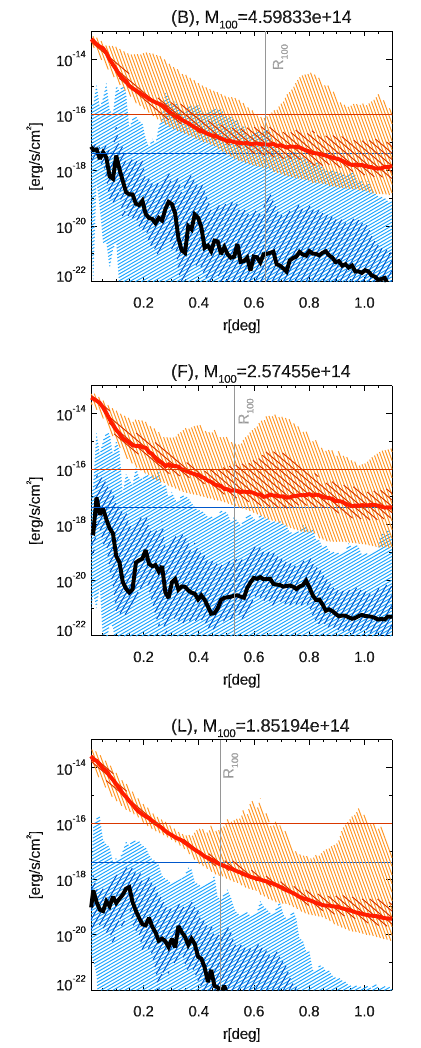}
  \includegraphics[width=0.24\textwidth]{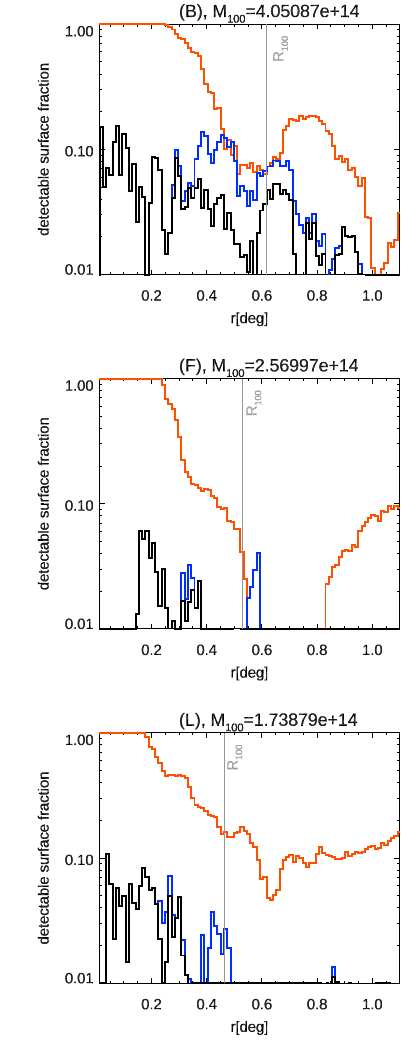}
\caption{Left panels: 2D X-ray (in the [0.8-1.2] keV range, in red/orange curves) and radio emission (at 260 MHz, blue curves) profiles for the 3  halos in our volume at $z=0.05$, showing the median (thick lines) and the $33-66\%$ and $5-95\%$ percentile ranges (shaded areas). The horizontal axis gives the angular distance from the centre of each cluster. The horizontal lines gives the reference X-ray (red) and the radio (blue) detection threshold considered in the paper, while the vertical grey lines give the location of $R_{100}$ for each cluster. Right panels: radial profiles of the surface fraction that can be detected in X-ray (red), in radio (blue) or by both (black) for the same objects.}
\label{fig:cluster_prof}
\end{figure}

Next, we analyse the distribution of X-ray and radio emission around our  clusters, beginning with the projected radial profile of their emission.  In Fig.~\ref{fig:cluster_prof}, we show radial distributions of three relevant examples for the average X-ray ([0.8-1.2] keV band) and radio (260 MHz) emission in 2D shells around their respective X-ray peaks. For each object, we show the median profiles (thick lines) as well as the $5-25-75-95\%$ regions around the median. The choice of cutting the profiles for $\geq 95\%$ limits the contribution from X-ray bright gas clumps, which get typically masked in real observations, and are not a good tracers for the ICM/WHIM conditions \citep[e.g.][]{ro06,zhur11,2013A&A...551A..22E}. The surface fraction that can be detected in the X-ray or in the radio domain as a function of the angular distance from the cluster centre of the 20 most massive clusters in our sample is given in the right panels of the same figure.  For each cluster, the vertical axis gives the fraction of the surface that is above the detection threshold in X-ray, in radio, or simultaneously in both. 

In roughly half of our sample a significant level of the X-ray emission should be detectable by \athena\ at the projected $R_{\rm 100}$ (consistent with Fig.\ref{fig:M100_erg}, which typically corresponds to $\sim 0.4-0.6$ degrees from the cluster centre. Conversely, in the radio domain several clusters host detectable radio emission in their innermost regions, which is associated to transient powerful merger shocks, leading to "classic" radio relics \citep[][]{2019arXiv190104496V}. Even with the planned sensitivity of SKA-LOW at $260$ MHz, the average radio emission there will be well below the detection threshold in most cases, and only in $\sim 10-20 \%$ of clusters we expect direct single detection of the shocked WHIM at the virial radius {\footnote{We notice that here our emission model is more conservative with respect to our previous analysis in \citet{va17cqg}, in which a 10 times larger initial seed field was assumed.}}, while a detection is guaranteed via stacking techniques \citep[][]{va17cqg}.

The second and third clusters shown in Fig.~\ref{fig:cluster_prof} (identified as "F" and "L" in the panels) are most frequent cases: there a non-negligible fraction ($5-20\%$) of the 
 $R_{\rm 100}$ shell is detectable in X-rays with {\athena}, while in both cases only a very tiny fraction of radio emission would be detectable by SKA-LOW. Moreover, the is no spatial overlap between X-ray and radio detection, hence the {\athena} and SKA will not be able to study the same portion of the ICM/WHIM in these objects, which are the majority in our sample.
 
 However, in the first system ("B" in Fig.\ref{fig:cluster_prof}) a $\sim 10\%$ of the $R_{\rm 100}$ shell will be detectable in X-ray with {\athena}, as well as a $\sim 5-10\%$ in the radio domain, with a non-negligible overlap of detectable regions in the two instruments, offering the chance joint analysis. A striking difference emerges from the X-ray profile, and may offer the key to predict the occurrence of such rare configurations: there is a prominent secondary X-ray emission peak outside of the virial radius of the first cluster, indicating the presence of a massive nearby cluster companion. This is not present in the profile of cluster "L", while a secondary peak is seen in cluster "F" but at a much larger distance from the virial radius of the main cluster. 
 
 Systems such as "B" only account for $\sim 10-20\%$ of our sample, and they suggest that {\it the presence of a massive companion, likely in an early interaction stage, is key to producing X-ray detectable bridges that are engulfed by detectable radio emission. How much general is this finding?}
 
\begin{figure}
  \includegraphics[width=0.45\textwidth]{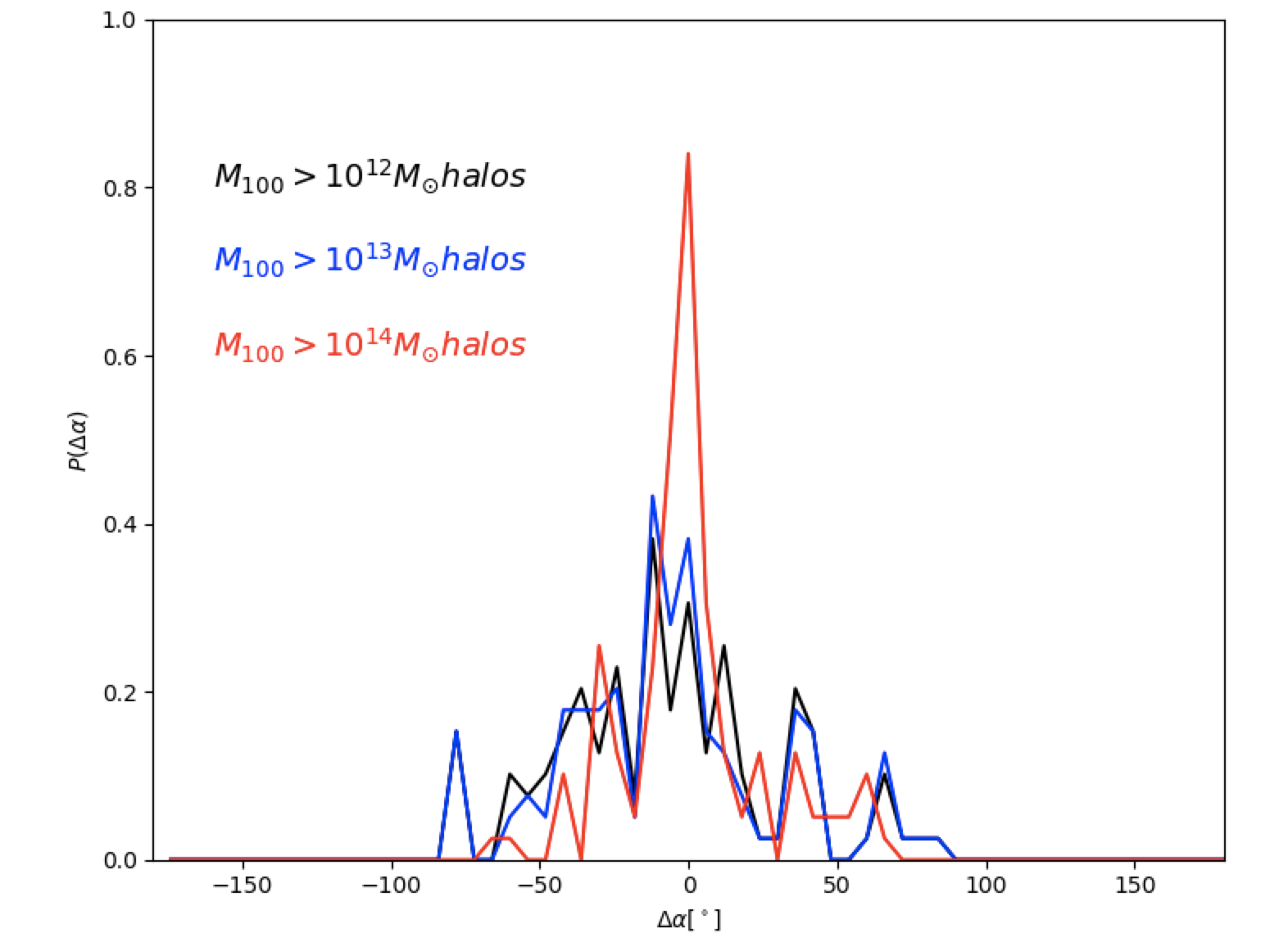}
\caption{Distribution of position angle between double detectable pixels at $z=0.05$ and the two closest galaxy clusters to each of them (see text for more explanations). The different colors refer to different cuts in the cluster mass within our sample.}
\label{fig:delta_alfa}
\end{figure}

 We further explored the link between joint X-ray and radio detections and merging clusters by studying the distribution of relative position angles between cluster pairs found around doubly detectable pixels in our sky model.
In Fig.~\ref{fig:delta_alfa}, we show the distribution of $\Delta \alpha$, the angle between the projected lines connecting the centres of two most massive clusters (considering different mass cuts) around doubly detectable pixel. 
If the two lines connecting the centres of each cluster are perfectly aligned, $\Delta \alpha = 0^\circ$, while large values of  $|\Delta \alpha|$ indicate a misalignment between detectable pixels and surrounding pairs of clusters. The distribution of angles peaks at $\Delta \alpha \sim 0^{\circ}$ and most cluster pairs (especially when $\geq 10^{14} M_{\odot}$) have $|\Delta \alpha| \leq 25^{\circ}$.\\

\begin{figure}
  \includegraphics[width=0.45\textwidth]{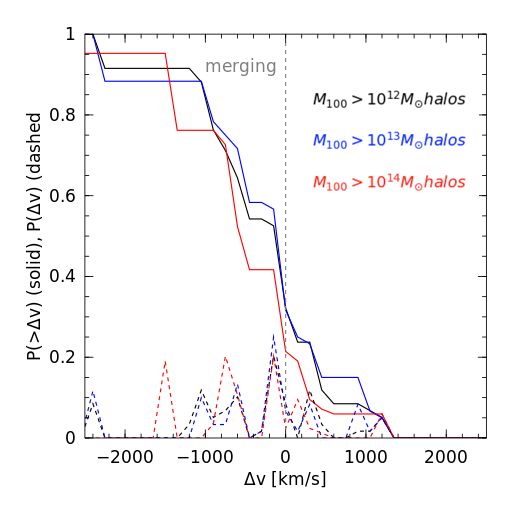}
\caption{Cumulative (solid) and differential (dashed) distributions of the relative velocities between the two closest galaxy clusters around all doubly detectable pixels in our run at $z=0.05$, limited to objects with a position angle $|\Delta \alpha| \leq 25^{\circ}$ in Fig.\ref{fig:delta_alfa}.  The different colors refer to different cuts in the cluster mass within our sample.}
\label{fig:delta_vel}
\end{figure}

We can also measure the relative gas velocities (measured as the median in the core cluster region) between clusters with $|\Delta \alpha| \leq 25^\circ$, for the same mass selections. This is shown in Fig.\ref{fig:delta_vel}, where we computed the (cumulative and differential) distributions of these relative velocities.
Especially when clusters with masses larger than $10^{14} \rm M_{\odot}$ are considered, in $\sim 80\%$ of cases their relative velocities  between the cluster cores are significantly negative, $\Delta v \sim - (1000 \div 3000) \rm ~km/s$. This confirms that most of such systems are colliding,  approximately along the straight line connecting their cores, where also most of our doubly detectable pixels are located.

Such cluster-cluster bridges are remnants of large-scale filaments once connecting these systems.  By the time at which interacting systems have nearly intersecting virial regions, the gas flow in between them   reaches  $\sim 1000-2000 ~\rm km/s$ velocities, and this forces their outer gas layers to be significantly compressed. 
 This compression causes a boost of the X-ray emission level, roughly of order $\sim (\rho_2/\rho_1)^{11/4}$ if $\rho_2$ is the gas density after the compression and $\rho_1$ was the gas density before the close encounter  \footnote{This follows from considering $L_X \propto \rho^2 \sqrt{T}$ and $T \propto \rho^{3/2}$ in the adiabatic case.}. Moreover, the increase in gas temperature boosts the chance of detection this tenuous gas phase in the [0.8-1.2] keV, by letting a larger portion of the X-ray emission spectra to "enter" the soft X-ray window. 
On the other hand, the  implied shock velocity are modest, with typical Mach numbers of ${\mathcal M} \leq 5-10$, fragmented on $\sim 10^2$ kpc scales. We will analyze in detail one of such systems at higher spatial resolution in Sec.\ref{subsubsec:sixte}. 

Such doubly  regions are parts of the WHIM that resemble the periphery of clusters. As already noticed by other authors \citep[e.g.][]{ro06,iapichino11} the usage of a too rigid temperature selection to define the WHIM phase, e.g. $T \sim 10^5 - ~10^{7} ~\rm K$, does not fully capture the different possible histories of the weakly X-ray emitting gas that may be detected in the periphery of galaxy clusters.

 \begin{figure*}
  \includegraphics[width=0.495\textwidth]{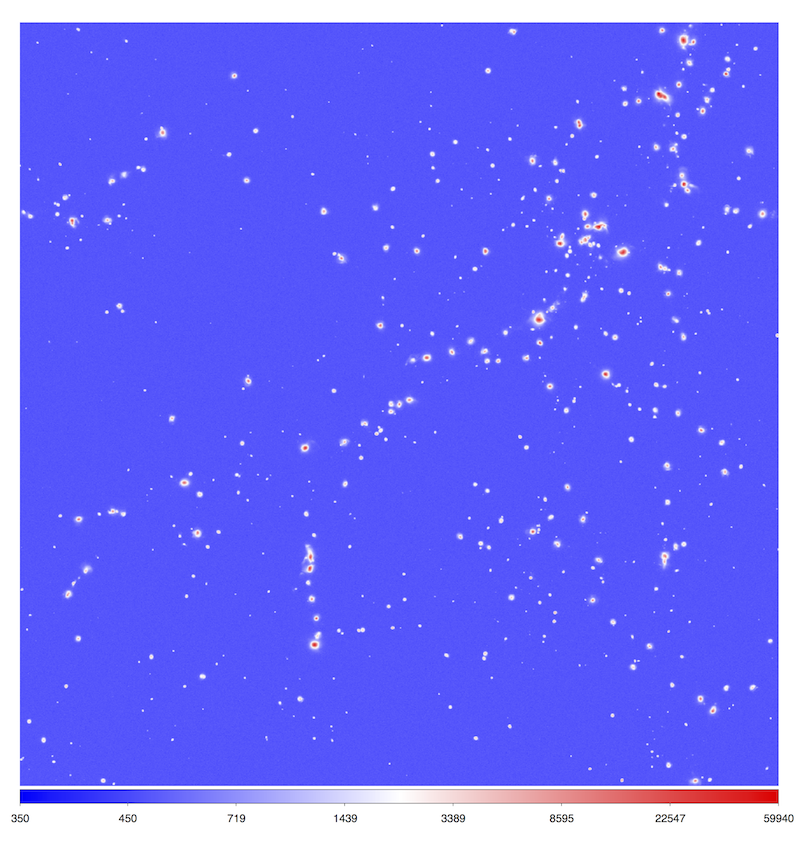}
    \includegraphics[width=0.495\textwidth]{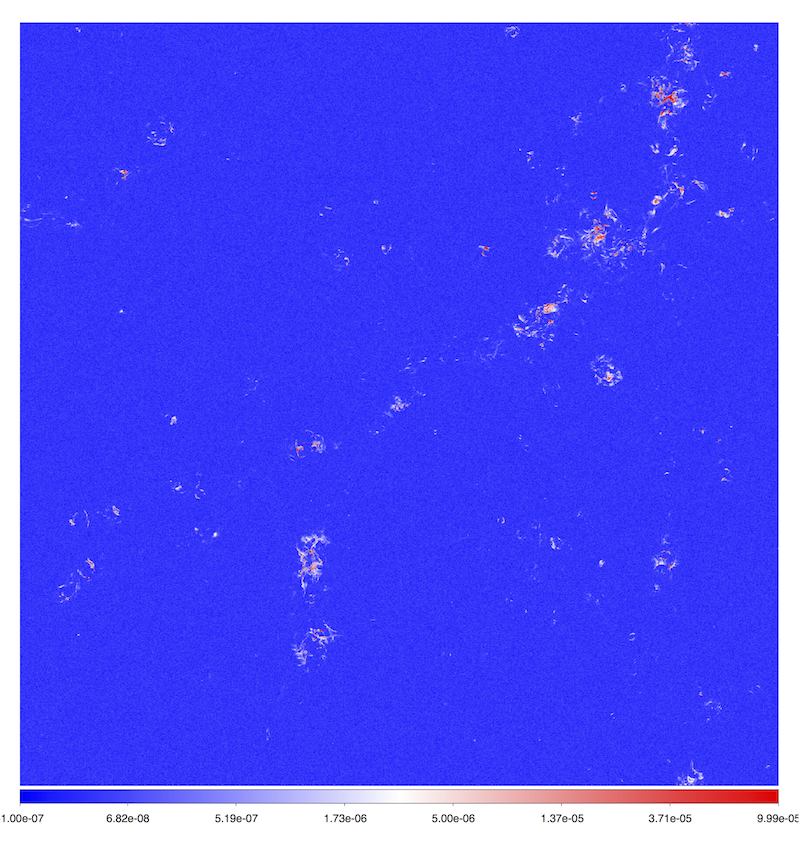}
\caption{Mock X-ray (left) and radio (right) observations of 
 the same simulated box of Fig.~1, located at $z \approx 0.05$. The left panel shows the photon counts for a $1 ~\rm Ms$ integration in the [0.8-1.2] keV band using WFI (see Sec.~\ref{subsubsec:xray} for details) while the right panel 
 gives the result of a mock radio survey at $260 ~\rm MHz$ with the SKA-LOW (see Sec.~\ref{subsubsec:radio} for details), in units of [$\rm Jy/arcsec^2$]. Each panel covers $\approx 28.4^{\circ} \times 28.4^{\circ}$in the sky. Both images include the noise level expected for the respective instrument, band and integration time (in the case of SKA-LOW, we also consider the confusion noise level).} 
\label{fig2}
\end{figure*}

 \begin{figure*}
  \includegraphics[width=0.495\textwidth]{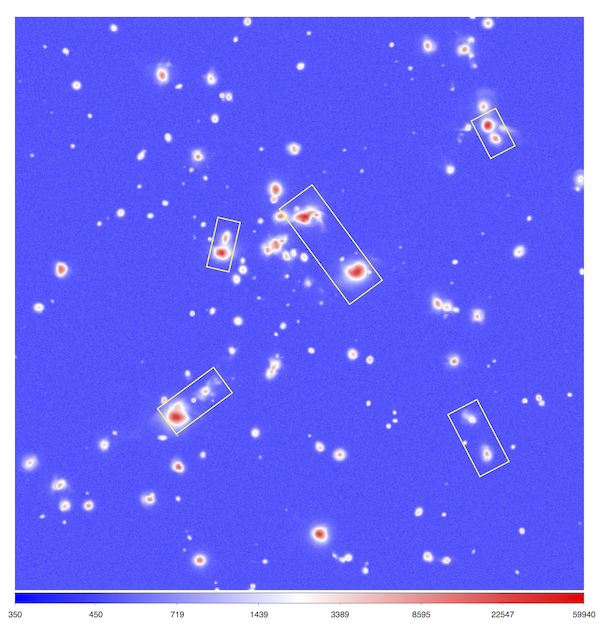}
    \includegraphics[width=0.495\textwidth]{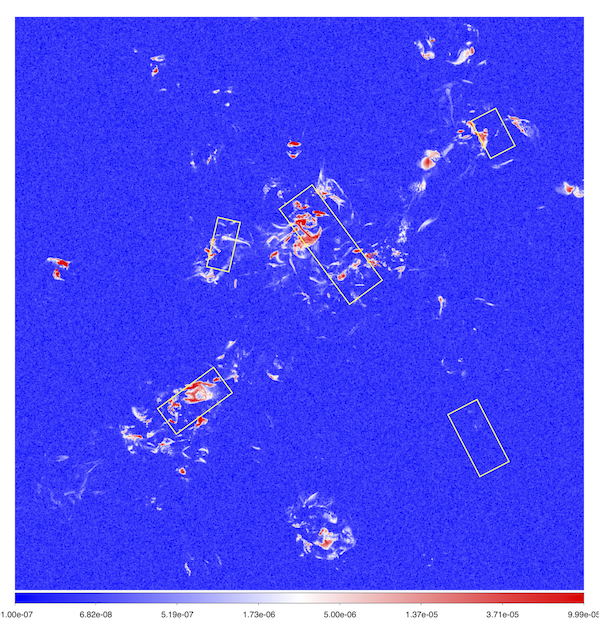}
\caption{Close-up view of a crowded $\approx 30 \times 30 ~\rm Mpc^2$ region of Fig.~2.  Each panel covers $\approx 8.5^{\circ} \times 8.5^{\circ}$in the sky. As in Fig.\ref{fig2}, the left panel gives the photon counts for a $1 ~\rm Ms$ integration in the [0.8-1.2] keV band using WFI while the right panel 
 gives the result of a mock radio survey at $260 ~\rm MHz$ with the SKA-LOW, in both cases with noise included. The additional yellow rectangles show pair of clusters with ongoing mergers.}
\label{fig3}
\end{figure*}

\begin{table}
\caption{Values adopted for our mock X-ray observations with {\athena}, eROSITA and XMM-Newton.
For each different energy band we give the count rate due to the effective (sky+instrumental) background, 
the fraction $f_{\rm abs}$ of source counts un-absorbed by the galactic column density 
(assuming  $n_H=2 \cdot 10^{20}\rm cm^{-2}$), 
and the mean effective collecting area $A_{\rm eff}$ in that energy range. 
For eROSITA, we quote values extrapolated from  http://www.mpe.mpg.de/1303215/eROSITA\_background\_v12.pdf (Boller, priv comm.); 
for XMM-Newton, we quote an average value on a set of observations obtained under different conditions
(of orbits, solar cycle, time; Ghirardini, priv. comm.).}
\centering \tabcolsep 5pt 
\begin{tabular}{c|c|c|c|c}
Instrument & Energy Band & $B_{\rm bg}$ &   $f_{\rm abs}$  & $A_{\rm eff}$\\  
& $[\rm keV]$   &   $\frac{\rm counts}{\rm arcmin^2 Msec}$  &   & [$\rm cm^2$] \\  
&  &     &   &  \\\hline 
 {\athena}-WFI & 0.3-0.8 & $2.1 \cdot 10^4$  & 0.83 & 9511 \\
  & 0.8-1.2 &  $3.1 \cdot 10^3$  & 0.95 & 12139 \\
  & 1.2-2.0 &  $1.4 \cdot 10^3$  & 0.98 & 10841 \\
  & 2.0-5.0 &  $3.4 \cdot 10^3$  & 0.99 & 4673 \\
  & 5.0-7.0 &  $2.0 \cdot 10^3$  & 1.00 & 2131 \\ \hline
  
  eROSITA & 0.3-0.8 &  $2.2 \cdot 10^3$  & 0.83 & 610 \\
  & 0.8-1.2 &  $4.6 \cdot 10^2$  & 0.95 & 1243\\
  & 1.2-2.0 &  $4.2 \cdot 10^2$  & 0.98 & 1267 \\
  & 2.0-5.0 &  $4.1 \cdot 10^2$  & 0.99 & 287 \\
  & 5.0-7.0 &  $3.0 \cdot 10^2$  & 1.00 & 88 \\ \hline
  
  XMM & 0.3-0.8 &  $4.6 \cdot 10^3$  & 0.83 & 1056 \\
  (PN + 2MOS)&0.8-1.2 &  $1.4 \cdot 10^3$  & 0.95 & 1655 \\
  & 1.2-2.0 &  $1.8 \cdot 10^3$  & 0.98 & 1894 \\
  & 2.0-5.0 &  $3.4 \cdot 10^3$  & 0.99 & 1337 \\
  & 5.0-7.0 &  $1.7 \cdot 10^3$  & 1.00 & 998 \\ \hline
\end{tabular}
\label{tab_bkg}
\end{table}

\subsection{Synthetic observations}
\label{mock}

Here we discuss strategies to detect the gas emission located in bridges of close cluster pairs. 
In order to quantify the detectability of intracluster gas bridges with existing or future observations, we 
produced a small survey of mock observations in X-ray and radio bands, assuming either the specifications for wide-area surveys of specific instruments, as well as of dedicated reprocessing (e.g. with UV tapering in the radio case) of specific objects in order to increase the sensitivity to diffuse emission.

\subsubsection{X-ray:  \athena\ , eROSITA and XMM} 


Here we compared the performances of shallow full-sky surveys with eROSITA (either 1.6 ks in the entire sky or 20 ks in the polar regions), with long targeted exposures with {\athena}'s Wide-Field Imager (1 Ms or 100 ks) and with XMM-Newton (100 ks), for different energy bands. 

To quantify the observable regions of each mock observation we use the signal-to-noise ratio (S/N), quantified as: 
\begin{equation}
S/N =   \frac{f_{\rm abs} \cdot S}{\sqrt {f_{\rm abs}(S+2B_{\rm bg})}}
\end{equation}
where $S$ is the number of photon counts originating from the source within a given energy band and collected by the effective area $A_{\rm eff}$,  $f_{\rm abs}$ is the ratio between the absorbed (from an assumed galactic column density 
of $n_{\rm H}=2 \cdot 10^{20} \rm cm^{-2}$) and not absorbed source counts for a typical plasma with gas temperature
of 3 keV, metallicity of 0.3 times the solar value,
and $B_{\rm bg}$ is the estimated total X-ray background, including the contributions from the Milky Way, 
the unresolved (20\% of the total) cosmic X-ray background and the instrumental background
{\footnote{
We use the responses available for each telescope at the following addresses:
https://www.the-athena-x-ray-observatory.eu/resources/simulation-tools.html (for \athena), http://www2011.mpe.mpg.de/eROSITA/response/ (eROSITA), 
https://www.cosmos.esa.int/web/xmm-newton/epic-response-files (XMM-Newton).}}. 
The details of the mock observing parameters considered here  are given in Tab.~\ref{tab_bkg}.

The left panels of Fig.~\ref{fig2}-\ref{fig3} show the mock X-ray exposure maps of our  simulated $100^3 ~\rm Mpc^3$  located at $z=0.05$ in the [0.8-1.2] keV band, for a 1 Ms exposure in every pixel of the box and considering  background counts as in Table~\ref{tab_bkg} for an {\athena}-WFI observation. The yellow rectangles in Fig.\ref{fig3} show that extended tails of detectable X-ray emission can often be seen in between closely interacting systems (only a fraction of which corresponds to physically bound objects actually interacting in 3 dimensions). 

To quantify the fraction of the gas in the cosmic web which can be detected by each instrument in different bands, it is convenient to measure the typical signal-to-noise that each observing strategy can achieve as a function of the local temperature, which we can parametrise through the mass-weighted mean temperature along the line of sight within each pixel. 

\begin{figure}
  \includegraphics[width=0.45\textwidth]{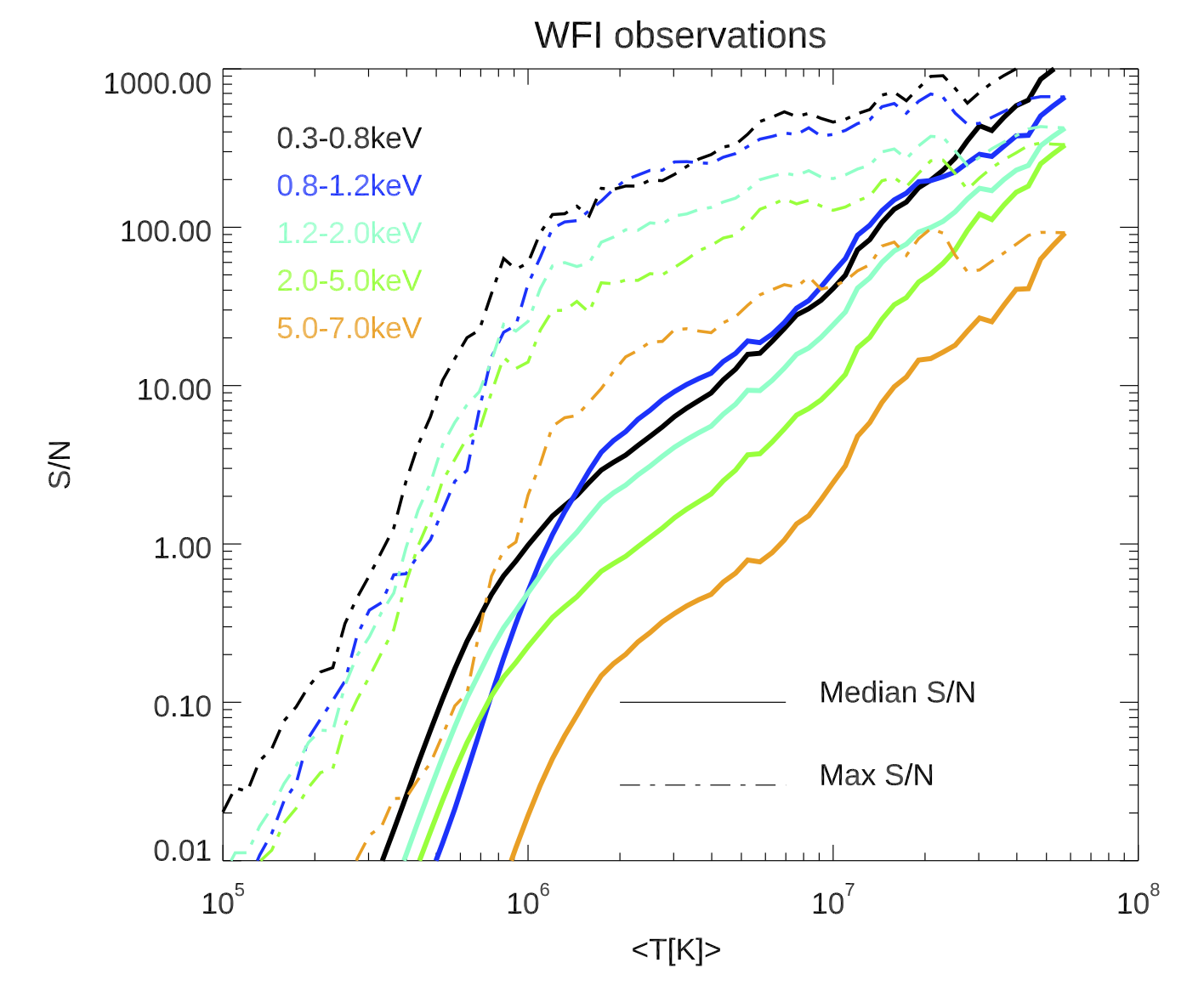}
  \includegraphics[width=0.45\textwidth]{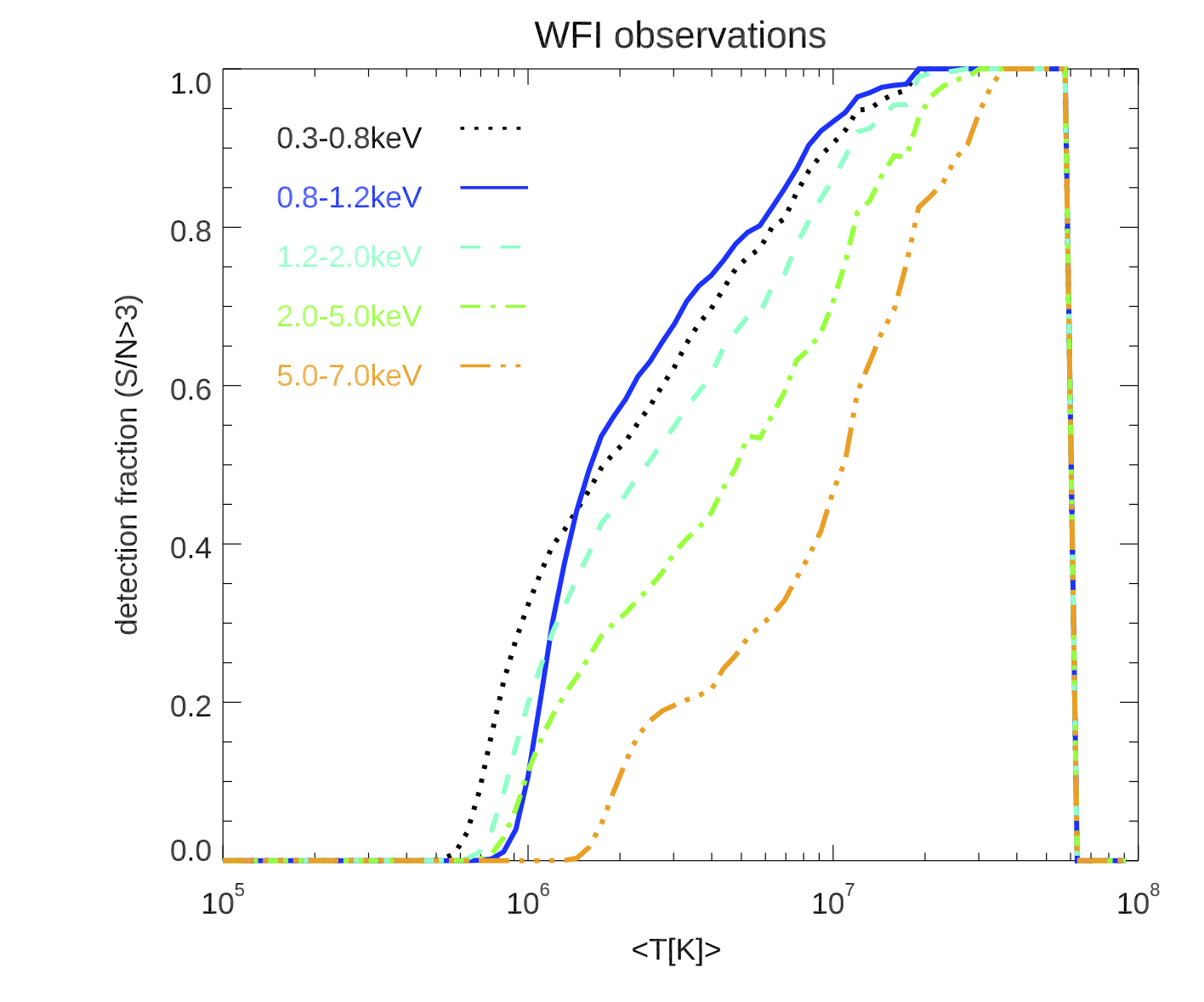}
   \caption{Top panel: median and maximum $S/N$ of  pixels in our simulated WFI observation, assuming a 1Ms integration, as a function of environment and for different energy ranges. Bottom panel: fraction of pixels with $S/N \geq 3$ as function of temperature and energy bands for the same mock {\athena}-WFI observations. }
\label{fig:X-ray_bands2}
\end{figure}

Fig.~\ref{fig:X-ray_bands2} displays the median/maximum $S/N$ for pixels in different environment and for different energy ranges in {\athena}'s WFI integration of 1 Ms.  The lower panel in the same Figure shows the detectable fraction of the sky model (marked as $S/N \geq 3$) as a function of temperature. As anticipated in Sec.~3.1.1, the combination of the instrumental and particle background, of the galactic absorption and of the effective collecting area as a function of energy makes the [0.8-1.2]keV slightly better compared to the [0.3-0.8]keV range for a detection of the emission from the baryons in the (projected) gas temperature range $T \geq 10^6-10^7 \rm ~K$,  which are more connected to cluster outskirts and cluster-cluster bridges.
In this range, we expect that $\sim 40\%$ ($\sim 80\%$) of the gas with projected gas temperature of $\sim 10^6 \rm ~K$ ($\sim 5 \cdot 10^6 \rm ~K$) can be detected with high significance with a 1Ms integration, respectively.
The detection fraction drops
to $\sim 20\%$ at $T \sim 10^6 \rm ~K$ in the [2.0-5.0] and to $\leq 1\%$ in the [5.0-7.0]keV range, respectively.

\begin{figure}
  \includegraphics[width=0.495\textwidth]{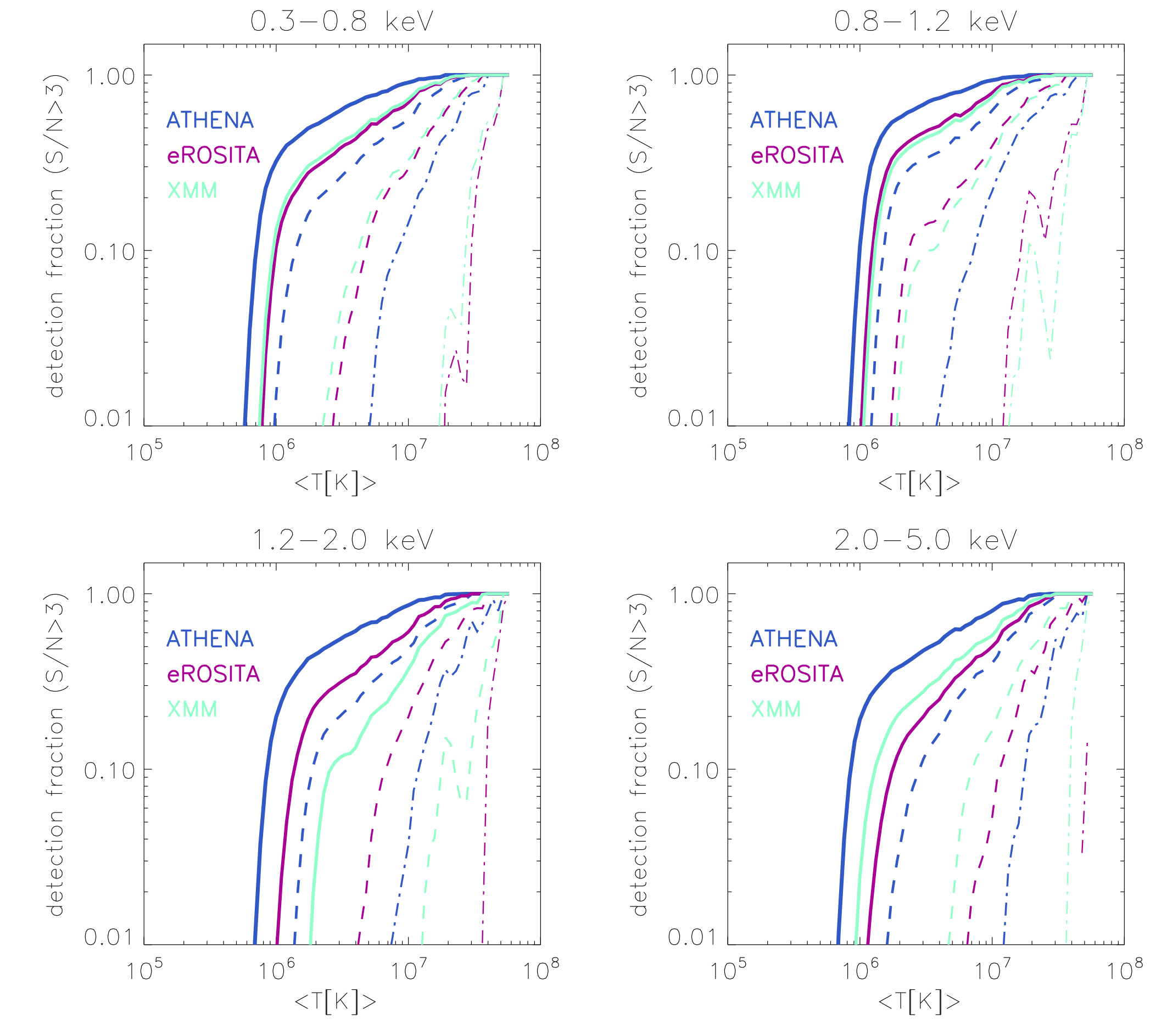}
\caption{Distribution of the detectable fraction ($S/N \geq 3$) of the simulated cosmic web as a function of the projected mass-weighted gas temperature, for the different X-ray satellites considered in this work, for different energy ranges and considering a 10 ks (dot-dashed lines), a 100 ks (dashed) and 1Ms (solid) integration. }
\label{fig:Xray_exp}
\end{figure}

In Fig.\ref{fig:Xray_exp} we compare instead the distribution of the detectable fraction of the sky model 
for {\athena}, eROSITA and XMM, for hypothetical 10 ks (dot-dash lines), 100 ks (dashed) or 1 Ms (solid) exposures. The better performances expected from {\athena} across the entire energy range stem for the large collecting area in each energy bin.
Owing to their smaller $A_{\rm eff}$, for the same exposures XMM and eROSITA yield significantly lower (factor $\sim 2-3$) detection fraction for $\leq 10^{7} \rm K$ in the [0.3-0.8]keV and [0.8-1.2]keV bands, while their performances drop more ($\sim 10$) in the  $\geq 1.2$ keV energy range. 
Below $1.2$ keV, our statistics suggest that $\sim 1$Ms integrations with XMM and eROSITA would be competitive with a $\sim 100$ ks intergation with ATHENA. 
However, in practice eROSITA is designed to perform shallower all-sky surveys \citep[e.g.][]{2014A&A...567A..65B}, integrating for $\approx 1.6$ ks in most of the sky, and up to $\approx 20$ ks for the regions around the poles. 

On the other hand, $\sim 50$ ks exposures with XMM have been performed in the outer regions of galaxy clusters, leading for example to the detection of intracluster filaments in A2774  \citep[e.g.][]{2015Natur.528..105E}.

\begin{table}
\caption{Assumed values for the radio observing parameters considered in this work: central observing frequency, beam resolution, thermal rms noise per beam and detection threshold considered in our analysis (considering a $3 \sigma$ detection, also including confusion noise) .}
\centering \tabcolsep 5pt 
\begin{tabular}{c|c|c|c|c}
Telescope & Frequency & beam & $\sigma_{\rm rms}$ & detection thr. \\  
 & $[\rm MHz]$  &  ["] & $[\rm \mu Jy/beam]$ & $[\rm \mu Jy/arcsec^2]$ \\\hline 
SKA-LOW & 260 &  7.3  & 4.8 & 0.24\\
LOFAR-HBA & 120 &   25  &   250& 1.05  \\
MWA Phase I & 200 & 120& 10,000 & 1.83\\
\label{tab_radio}
\end{tabular}
\end{table}

\subsubsection{Radio: SKA-LOW and LOFAR}

We also produced mock radio observations at a given frequency $\nu$, for the all-sky survey planned for  MWA, LOFAR-HBA and  SKA-LOW with a procedure similar to \citet{va15radio} and \citet{va15ska}.
The radio sky models were transformed using Fast Fourier Transform to remove the frequencies below the minimum antenna baseline of each specific radio configuration, i.e. we mimic the loss of signal from scales larger than those sampled by the minimum instrumental baseline. This is particularly relevant for the large-scale diffuse gas emission from filaments and cluster outskirts, even if all these low-frequency telescopes are suitable to well sample such large-scale fluctuations, as well as telescopes working at $\sim 1.4$ GHz. Our maps are converted back into real space and the emission is convolved for the resolution beam with a Gaussian filter, and the detectable emission is only that  $\geq 3 \sigma_{\rm rms}$ (where $\sigma_{\rm rms}$ is largest between the thermal or the confusion noise of each instrument). 
In this simplistic approach we assume that it is possible to entirely remove galactic foregrounds \citep[e.g.][]{2015MNRAS.447.1973B} and point-like radio sources \citep[e.g.][]{2016ApJS..223....2V}, so that the theoretical thermal/confusion noise \citep[e.g.][]{2019MNRAS.tmp..395L} of each instrument can be reached. \\

The details of the mock observing parameters considered in this work are given in Tab.\ref{tab_radio}, in which we compared the performances of our reference SKA-LOW  $260$ MHz observation to the lower $120$ MHz central frequency of LOFAR-HBA and to a $200$ MHz observation with MWA Phase I.  
The left panels of Fig.\ref{fig2}-\ref{fig3} show the mock radio observation of our  $100^3 ~\rm Mpc^3$ box at $z=0.05$ at the central observing frequency of $260 ~\rm MHz$, assuming the resolution, baseline sampling and sensitivity of the SKA-LOW observing conditions given in Tab.~\ref{tab_radio}.  The typical trends of detections we can expect from these instruments applied to the quest for the cosmic web have been already discussed at depth in previous work \citep[e.g.][]{va15radio,va17cqg} and are applicable also here, albeit for SKA-LOW we consider the most updated predictions on the survey performance.

While our main focus here is to assess the potential of each  all-sky radio survey in detecting promising candidates for longer X-ray integrations outside of the virial region of clusters, the peak performance on diffuse emission that each radio telescope can achieve are slightly larger, i.e. by observing the same target for longer exposures and/or tapering the data to a coarser resolution to increase the surface brightness sensitivity, if confusion noise is not the limiting factor. 
We comment on this issue more diffusely in the next Section.
In general, the conclusions reached in our previous works \citep[][]{va15ska,va15radio,va17cqg} apply also here: 
while the SKA-LOW promises to detect with higher fidelity a non-negligible fraction of the radio emitting cosmic web, both LOFAR and MWA should be able to perform earlier detections of cluster outskirts and emission in-between interacting clusters.

 \begin{figure*}
  \includegraphics[width=0.99\textwidth]{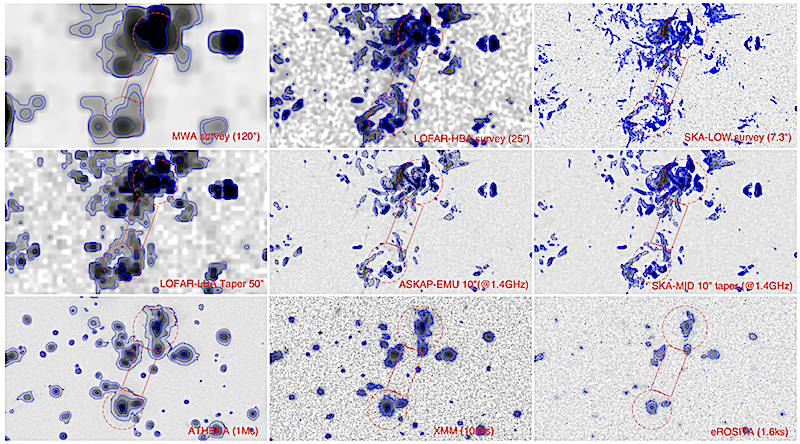}
\caption{Example of mock radio (top two rows) and X-ray (bottom row) observations of two bridges connected to a $\sim 2 \cdot 10^{14} M_{\odot}$ galaxy cluster in our simulation. The red circles denote the $R_{\rm 100}$ of the halos connected by bridges with width $5'$ (red rectangles), as in Sec.~3.3.3. In all panels, we give in colors the signal and noise of each mock observation, convolved for the resolution of each observation, while the blue contours are drawn with a logarithmic spacing, starting from $3 ~S/N$ of each observation (see text for explanations). }
\label{fig:bridge_map}
\end{figure*}

 \begin{figure}
  \includegraphics[width=0.495\textwidth]{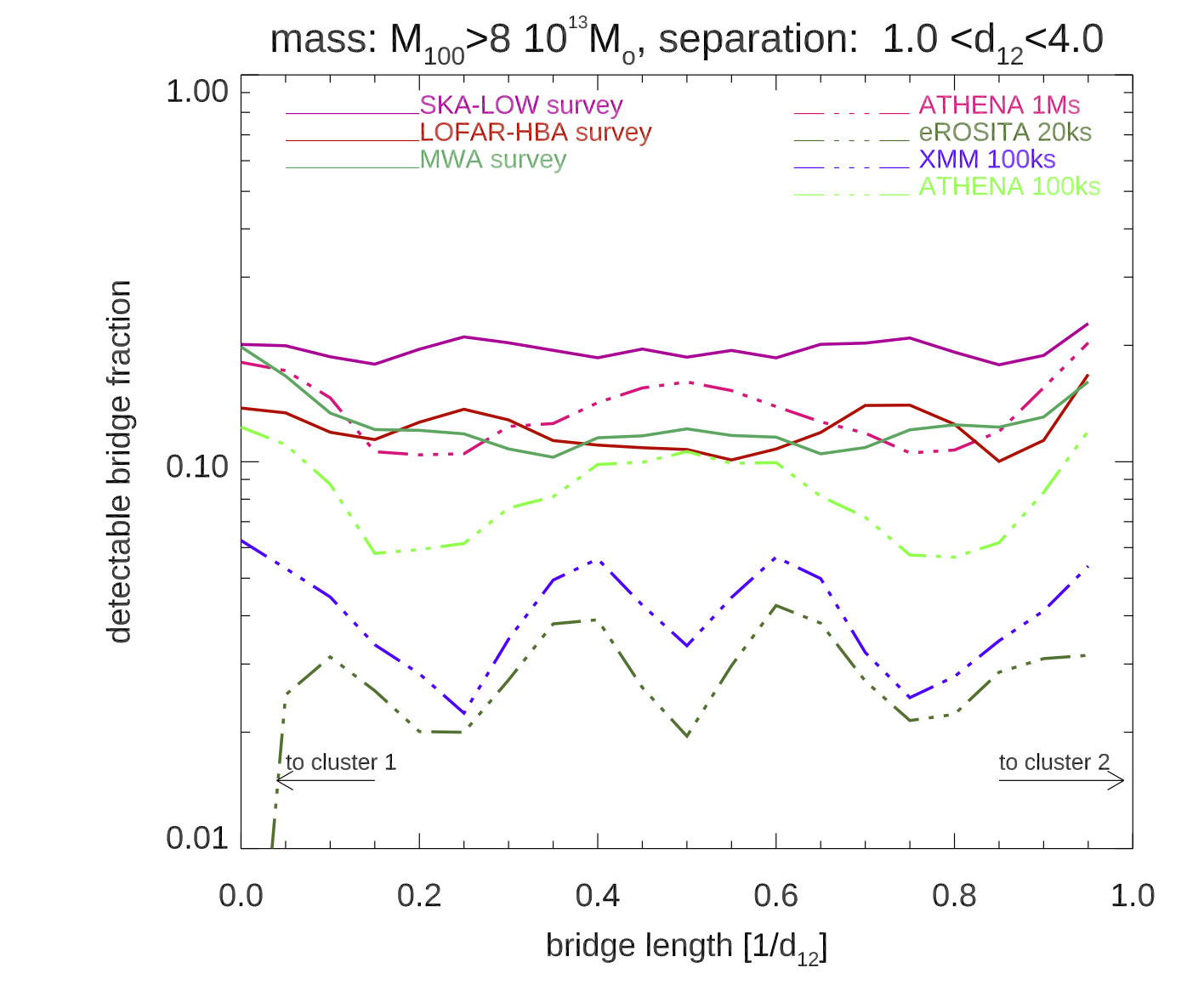}
\caption{1-dimensional profile of the detectable bridge  fraction connecting pairs of clusters with mass $M_{\rm 100} \geq 5 \cdot 10^{13} M_{\odot}$ , considering different  X-ray (solid lines) or radio (dot-dashed) observing configurations.}
\label{fig:bridge1}
\end{figure}

 \begin{figure}
  \includegraphics[width=0.495\textwidth]{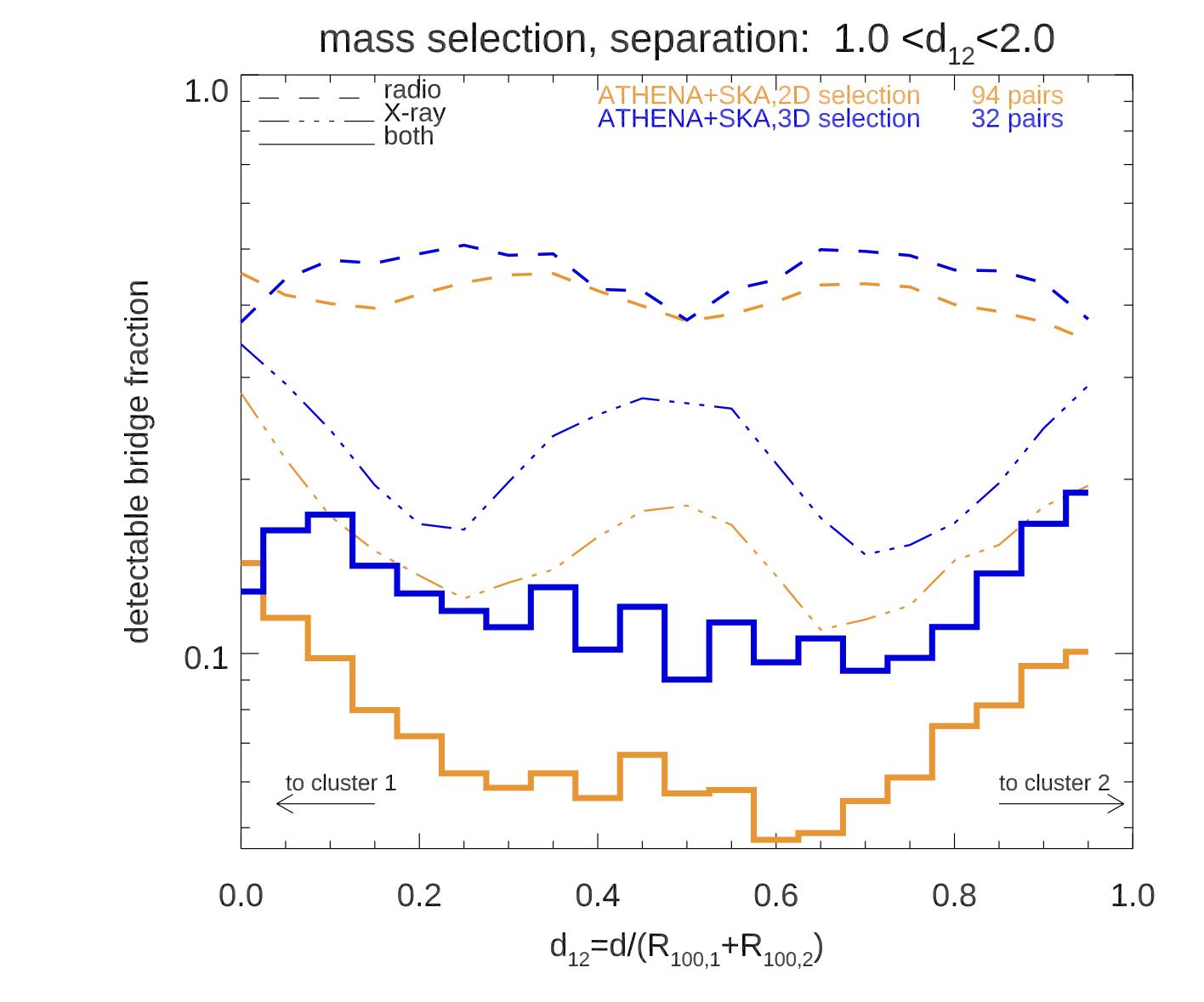}
\caption{Detectable fraction of bridges connecting pairs of clusters with mass $M_{\rm 100} \geq 10^{13} M_{\odot}$ and for {\athena} and SKA-LOW observations, for samples of cluster pairs selected only based on their projected (2D) separation, or including also a limit to their physical (3D) separation.}
\label{fig:bridge2}
\end{figure}

 \begin{figure}
  \includegraphics[width=0.495\textwidth]{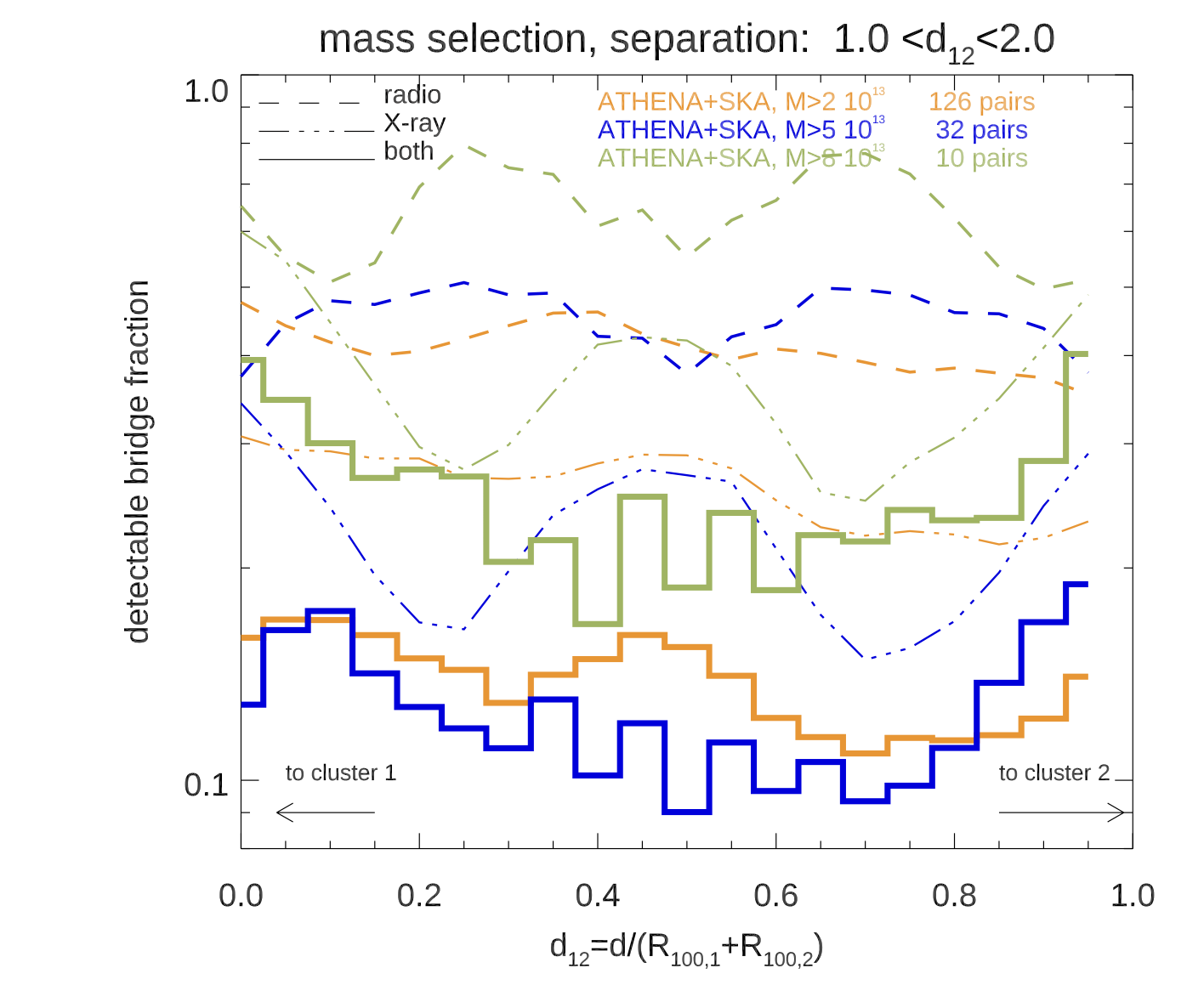}
  \caption{Detectable fraction of bridges connecting pairs of clusters with mass $M_{\rm 100} \geq 10^{13} M_{\odot}$ and for {\athena} and SKA-LOW observations, for samples of interacting clusters chosen after imposing an increasing lower bound for the minimum mass.}
\label{fig:bridge3}
\end{figure}

 \begin{figure}
  \includegraphics[width=0.495\textwidth]{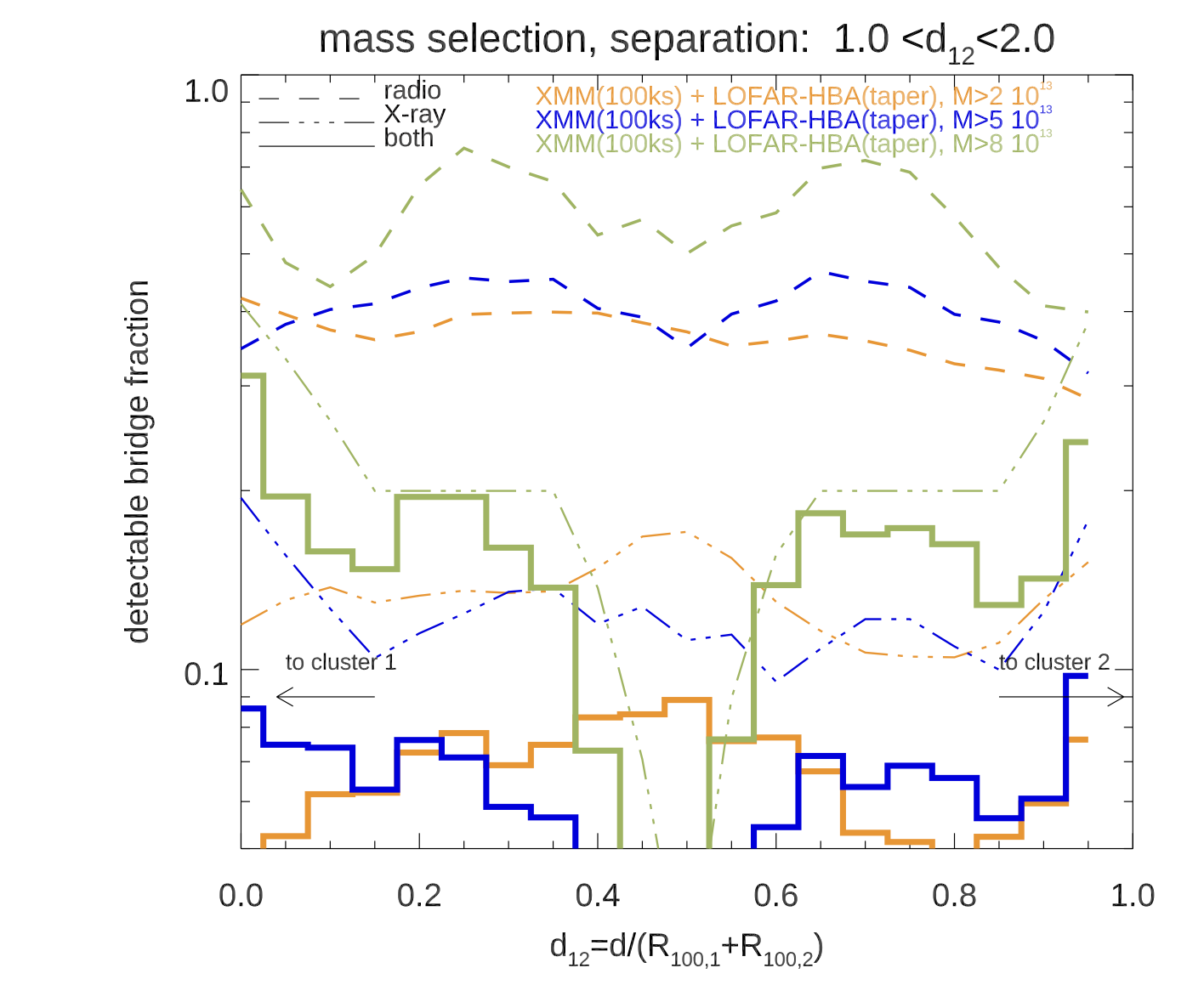}
  \caption{Detectable fraction of bridges connecting pairs of clusters with different masses, as in Fig.\ref{fig:bridge3}, but for 100ks exposures with XMM-Newton and LOFAR-HBA (50" taper) observations.}
 \label{fig:bridge4}
\end{figure}

\subsubsection{What can be detected with realistic observations?}

Simulations can help us in defining the most efficient strategy to select targets for long spectroscopic analysis of cluster bridges.

Based on our catalog of halos, we extracted all possible close cluster pairs, with the
priors that their virial spheres (based on $R_{\rm 100}$) should not overlap, and that their {\it projected} separation is smaller than four times $R_{\rm 12}=R_{\rm 100,1}+R_{\rm 100,2}$. 
We then select a rectangle along the line connecting their centres, starting from one virial radius and ending onto the other, with a  $5'$ width, which corresponds to {\athena}'s X-IFU field-of-view.

An example is shown in Fig.~\ref{fig:bridge_map}, where we show mock radio and X-ray observations of two bridges connected to a $\sim 2 \cdot 10^{14} M_{\odot}$ galaxy cluster.  Contours indicate the regions that can be detected at $\geq 3 \sigma$ in the different frequencies. 

All radio surveys should be able to detect emission from shocked gas in the larger bridge in the center of the image, while only SKA-LOW should detect some small emission patches in the smaller bridge. 
While the sensitivity in surface brightness can be increased by  changing the observing strategy, for example by ad-hoc tapering, e.g. \citet{2018MNRAS.478..885B}), the high spatial resolution of SKA-LOW ($\theta=7.3"$) is key to resolving the intricate shock network in bridges. In the same figure we also show that, unlike in the case of large-scale diffuse emission expected from cosmic filaments, the relatively small-scale ($\leq 5-10'$) emission from shocks in bridges is also well-sampled at higher frequencies. In this case, observations at $1.4$ GHz with ASKAP or SKA-MID could detect this emission.

Moreover, the possibility of observing polarised emission from such regions will enable SKA-MID to make more significant detections since polarisation reduces the dynamic range and the confusion level, which is the biggest constraint for SKA-LOW \citep[e.g.][]{va15ska}.

The lower row shows the performance of X-ray observations, considering a 1 Ms {\athena}-WFI integration of the same region, 100 ks with XMM or 1.6 ks as in the eROSITA all-sky survey.
Clearly, the detections of such regions in X-rays will remain a challenge, in which only {\athena} promises some joint detection with radio observations. {\footnote{Since the resolution of our simulation is coarser than the equivalent spatial resolution of these instruments at this redshift, the capabilities of {\athena} in resolving small-scale structures in the ICM volume are underestimated here.}}

In order to quantify the performance of real observations of intracluster bridges, we 
extracted from our catalog of pairs the fraction of the connecting rectangle ($5'$ wide) that can be significantly detected with various instruments and observing techniques. We then averaged the results by rescaling by the distance between the two virial spheres. With a few variations in the selection criteria and in the observing techniques, we find:

\begin{itemize}
\item {\it performance of X-ray vs radio observations:} in Fig.~\ref{fig:bridge1} we show the mean detectable fraction of intracluster bridges as a function of the distance from one virial radius to the other, and normalized for the projected separation between the two virial radii. We extracted here all pairs of clusters with 
$M_{\rm 100} \geq 8 \cdot 10^{13} M_{\odot}$ and found at a $\leq 4 R_{\rm 12}$ distance, both in projection and in 3D ($50$ pairs in total).  We consider here {\athena}-WFI 1Ms and 100 ks integrations, XMM 100 ks integration and eROSITA 20 ks integration, and observations with SKA-LOW, MWA Phase I and LOFAR-HBA surveys. On average, XMM and eROSITA integrations give a very little detection fraction (we also performed the analysis of a 1.6 ks survey with eROSITA, which yields virtually null detections in the entire sample, see e.g. the lower right panel of Fig.\ref{fig:bridge_map}). {\athena}'s 1Ms integration may instead detect a $\sim 10-15\%$ of the surface of intracluster bridges, but also a $100$ ks integration with the same telescope will detect a  significant $\sim 5-9\%$ of it. 
As expected, the situation is clearly better in the radio domain as an  SKA-LOW survey should be able to detect $\sim 20\%$ of the intracluster filaments area, and MWA Phase I and LOFAR-HBA surveys at least of $\sim 10-15\%$ of it. For radio surveys the detectable (area) fraction is not an unambiguous proxy for the real performance of each telescope, because the rectangular area we consider here is  $5'$ wide, which is close to the beam size of MWA Phase I observations ($2'$) or of what is typically achieved for tapered LOFAR-HBA observations ($50"$). Hence, a few detected beams can contribute to a large covering fraction, yet at the cost of a loss in spatial detail, which is crucial (see Fig.~\ref{fig:bridge_map}) if the emission is produced by several shock waves.
SKA-LOW will reach the maximum depth after only $\approx$ 10 hours of integration (due to the confusion limit), but it should be able to give a detailed spatial information on shocks within the bridge regions, which can also allow a detailed modelling of particle spectra and ageing.  

\item {\it separation between cluster pairs:} in Fig.~\ref{fig:bridge2} we show that if the only prior is on the projected separation (grey lines) a good fraction of the bridge is detectable with a 1 Ms integration with {\athena}. However, only $\sim 10-20\%$ is detectable with SKA-LOW, up to a distance of twice $d_{\rm 12}$. If we only consider clusters that are {\it physically} related, by imposing an additional prior ($d_{\rm 3D} \leq 10 ~d_{\rm 12}$) on their 3D distance, the chances of detections in radio, and hence of joint detections, increase very significantly and reach $\sim 15-20\%$ close to clusters. This follows from the fact that indeed the detectable radio emission is mostly produced by shocks associated with {\it physical} gas perturbations triggered during the early merger stage of such objects, rather than to the  simple superposition of the shocked outer layers of galaxy clusters which are not interacting;

\item{\it effect of mass selection:} in Fig.~\ref{fig:bridge3} we show the effect of using a higher mass threshold for the selection of our couples of objects (always connected also in 3D, using the previous prior). The fraction of the joint detection overall increases going to higher masses, even though the total number of objects goes down. For the highest mass selection, $\sim 20\%$ of the central bridge area (and $\sim 30\%$ closer to clusters) can be detected, both, by {\athena} and SKA-LOW.  We notice that if we repeat the same test assuming a 100ks survey with XMM-Newton, and a $50"$ tapered LOFAR-HBA observation, the trend with mass becomes less clear. This is driven by the overall lower rate of detections in X-ray makes this statistics dominated by small number statistics, as shown in Fig.\ref{fig:bridge4}. However, the performances of LOFAR-HBA tapered to this resolution, meant  to maximise the signal to noise to diffuse emission, are comparable with an SKA-LOW survey, as they give just a $\sim 25\%$ smaller fraction of detectable bridge area.
\end{itemize}

The limited volume of our simulation does not allow us to test a large number of cluster pairs. Hence our work can only identify global trends across the investigated cluster population. 
We conclude that the best strategy to detect X-ray emission from the WHIM in intracluster bridges is to select cluster pairs with short physical and projected separation ($1 \leq d/R_{\rm 12} \leq 4$, where $R_{\rm 12}=R_{\rm 100,1}+R_{\rm 100,2}$, with large masses ($M_{\rm 100} \geq 10^{14} M_{\odot}$) and with the additional prior of having detected radio emission between them. In this case, investing  $\sim 1$ Ms of observation time with {\athena} should result in a detection of the WHIM in emission from $\sim 20-30\%$ of the connecting area, which can in principle be suitable for spectroscopic analysis with X-IFU.

 \begin{figure*}
  \includegraphics[width=0.3\textwidth]{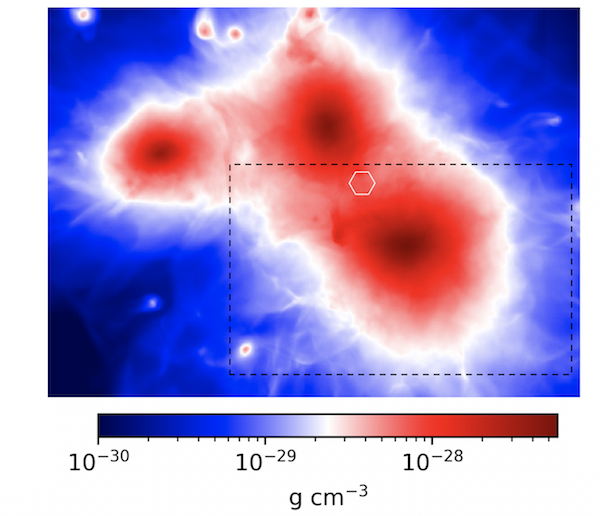}
  \includegraphics[width=0.66\textwidth]{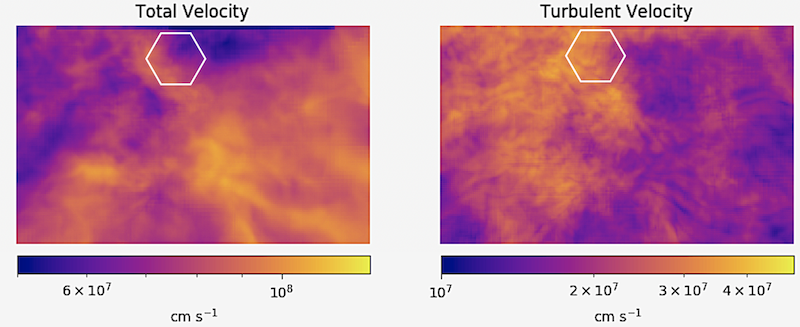}  
\caption{Left panel: projected mean gas density for the interacting cluster used for our SIXTE simulation in Sec.\ref{subsubsec:sixte}. Central and Right panels: projected (volume weighted) map of total gas velocity module and of turbulent velocity module for a zoomed region in the left panel. The additional white region shows the $5'\times 5'$ region used for our SIXTE mock observation. }
\label{fig_e5A}
\end{figure*}

\subsubsection{A pilot X-IFU observation of a gas bridge between interacting clusters}
\label{subsubsec:sixte}

Finally, we conducted a pilot study of a potentially promising target for future 1 Ms {\it Athena} X-IFU observations, applying the end-to-end X-IFU pipeline outlined in \citet[][]{roncarelli18}.

We performed the {\enzo} AMR re-simulation of a galaxy cluster in an early merger state at $z=0.1$, with a total mass of $M_{\rm 100}=6.37 \cdot 10^{14} M_{\odot}$ and a virial radius of $R_{\rm 100} = 2.05 ~\rm Mpc$. 

We selected this cluster from a different parent simulation, using nested initial conditions and 7 additional levels of AMR to increase the maximum spatial resolution to  $\Delta x_{\rm max}=3.95 ~\rm kpc$, as in \citet{va18mhd}. This cluster has been specifically selected as it shows at $z=0.1$ an ongoing merger with a second massive galaxy clusters (e.g. Dominguez-Fernandez et al. submitted), with indication of strong dynamical activity in the $R_{\rm 100}-R_{\rm 200}$ radial range from the first cluster. Moreover, by applying the same mock observation techniques introduced in the previous Sections, we identified in this object a patch, located in-between the two clusters, in which a significant detection will be possible both with SKA-LOW and with a long {\athena}'s integration. We also analysed this object using the multi-scale filtering technique to detect shocks and turbulent motions (and further decompose their velocity field into compressive and solenoidal components) as in \citet{va17turbo}.

As shown in Fig.~\ref{fig_e5A}, the region in between the two colliding clusters has a temperature of $T \sim 5-6$ $\rm keV$, significantly hotter than the surrounding ICM, as an effect of the ongoing merger. 
Compression rather than shocks is responsible for this heating, as the outer cluster shells of the interacting clusters get compressed at the velocity of $\sim 500-700$  $\rm km/s$ and the average density in the bridge is $\sim 10^2 \langle \rho_b \rangle$ (where $\langle \rho_b \rangle$ is the cosmic mean baryon density).  Due to the large temperature and sound speed, no strong accretion shocks are found. Instead, we find a network of $\mathcal{M} \sim 2-5$ shocks, intermittently driven by the event. Our turbulent filtering detects turbulent velocities of order $\sim 400-500 ~\rm km/s$ in this region, almost equally divided into solenoidal and compressive components,
with a typical outer scale of $\sim 400 \rm~ kpc$. The magnetic field in the bridge region where radio and X-ray emission can be detected is $B \sim 0.1-0.2 ~\rm \mu G$, which yields a plasma beta parameter of order $\beta_{\rm pl}=n k_b T/(B^2/8 \pi) \sim 5000$. 

From this region, we extracted a $(75\times75\times4800)$ kpc$^3$ box that contains the  entire emitting regions, as well as the outer regions of the two clusters in the field of view. The region's $z$-axis is aligned with the line-of-sight. Following \cite{roncarelli18}, we simulated with 1 eV resolution the X-ray emission assuming an \textsc{apec} model \citep[version 2.0.2,][]{smith01}, absorbed with a Galactic equivalent hydrogen column density $n_{\rm H}=2 \times 10^{20}$ atoms cm$^{-2}$. 
We assume a rather conservative value of (spatially constant) metal abundance $Z=0.2~ Z_{\odot}$ with respect to the solar abundance in \cite{ag89}. 
To provide a realistic observational setup, we included an X-ray background modelled by (i) two thermal local components with temperatures of 0.099 and 0.225 keV and normalization for the adopted \textsc{apec} model of 1.535 $\times 10^{-6}$ per arcmin$^2$ of integrated area, respectively, both with solar metallicity.  Only the second component is absorbed by $n_{\rm H}$ and with a normalization reduced by a factor of 0.42 \citep[see][]{mccammon02}. (ii) A cosmic X-ray background, assumed unresolved at the 20\% level, simulated with an absorbed power-law with a photon index of 1.5 and a normalization of $2.1 \times 10^{-7}$ photons/keV/cm$^2$/s/arcmin$^2$ at 1 keV \citep[see also, e.g.][]{lotti14}. 
After projecting these spectra into the plane of the sky, we used the resulting signal as an input for the SIXTE simulator \footnote{https://www.sternwarte.uni-erlangen.de/research/sixte/} \citep{schmid13}, assuming $ 1 ~\rm Ms$ exposure. 
An updated instrumental particle background \citep{lotti17} is already implemented in SIXTE.
The output of the SIXTE code consists of a realistic mock X-IFU observation, i.e. an observed event-list containing a total of $\sim 16500$ counts\footnote{As a further reference to the reader, we verified that $\sim$8900 counts are coming from the source (i.e. the simulated plasma), while $\sim 5400$ and $\sim 2200$ from the X-ray and particle background, respectively. This piece of information, that would not be available to the observer, has been omitted in the following analysis.} (0.3--10 keV) from a 0.382 arcmin$^2$ area in the center of the X-IFU field. 
These photons were folded in single spectrum that has been jointly fitted with a spectrum containing only the expected photons from a background modelled as described above, using a Cash statistic \citep{cash79} implemented in XSPEC \citep{arnaud96}. 
To derive the thermodynamic and kinematic properties of the gaseous component associated to the cluster, we assumed a \textsc{bapec} model with 5 free parameters, namely normalization, temperature, metal abundance, redshift and velocity dispersion ($n_{\rm H}$ has been fixed to the input value) and proceeded with a blind fitting analysis. 

In Fig.~\ref{fig_xifu}, we show the mock spectra and the fitting results, with the three components (thermal, X-ray and particle background). 
The results of the X-IFU spectral fitting simulation are also shown in Tab.~\ref{tab_z}, together with the reference values, computed as the averages on the simulation cells. 
Thanks to the 1 Msec exposure, temperature (compared to the spectroscopic-like) and metallicity are recovered with high precision with a typical statistical error of 0.3--0.4 keV and 0.05 $Z_{\odot}$, respectively. 
Most importantly, the evidence of a significant velocity dispersion is detected at high significance (more than $2.5 \sigma$), albeit with a relatively high statistical error. All the quantities are recovered with no apparent systematic bias. 

However, our setup assumes a perfectly known (X-ray and particle) background.
To relax this assumption, we have also run 100 Monte-Carlo spectral simulations assuming 
the cluster component and propagating random fluctuations, consistent with the current expectations on the background reproducibility in X-ray, on $n_{\rm H}$ (at 1\% level), on the particle background (2\%) 
and on the remaining parameters of the background model, i.e. the thermal components and the power-law for the unresolved cosmic X-ray background (5\%).
We estimate the following systematic scatter in the distribution on the best-fitting measurements of the 5 parameters of the \textsc{bapec} model: 
$\sim$ 2\%, 6\%, 27\%, 58\% and $<$1\% on normalization, temperature, metal abundance, line broadening and redshift, respectively.  
These results show that, while most of the parameters will be only limited by the photon counts statistic, the characterization 
of the emission lines in terms of total metallicity and broadening at this level of surface brightness will depend on a reliable modelization of the
underlying background.

\bigskip

\begin{figure*}
  \includegraphics[width=0.95\textwidth]{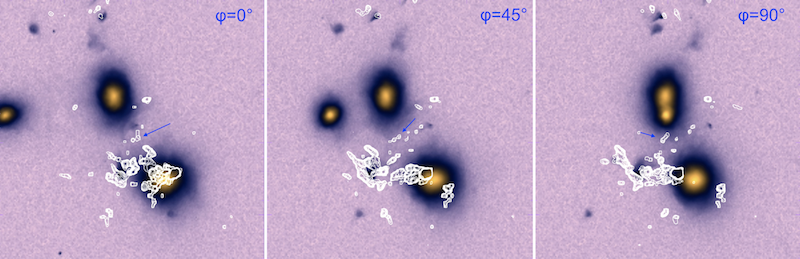}
\caption{Three different rotations ($0^\circ$, $45^\circ$ and $90^\circ$) around cluster E5A, showing the X-ray counts (colors) and the detectable radio emission at 260 MHz (contours). The additional blue arrow give the approximate location of the detectable emission region in the rotated images.}
\label{fig_rot}
\end{figure*}

\begin{figure*}
  \includegraphics[width=0.495\textwidth]{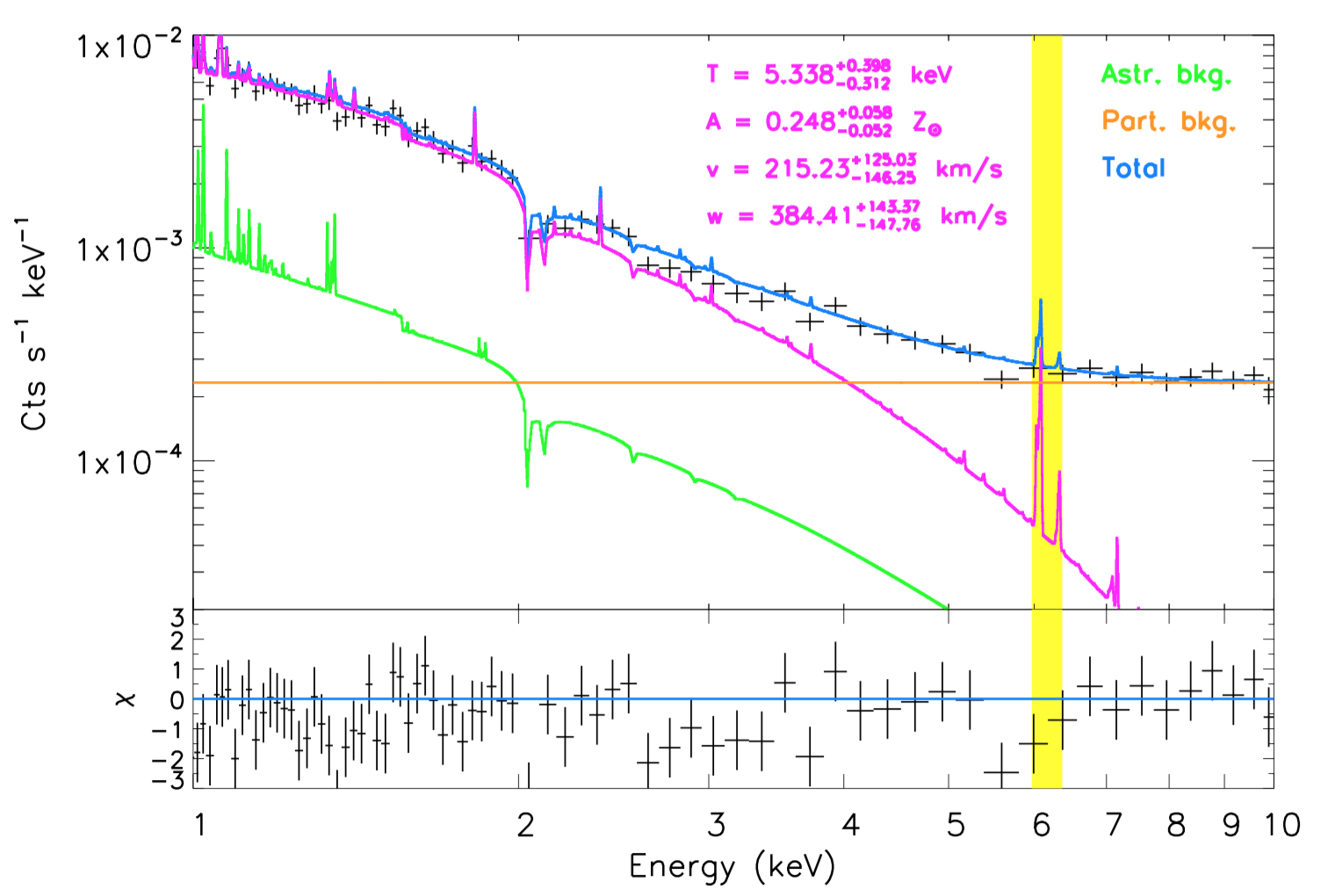}
 \includegraphics[width=0.495\textwidth]{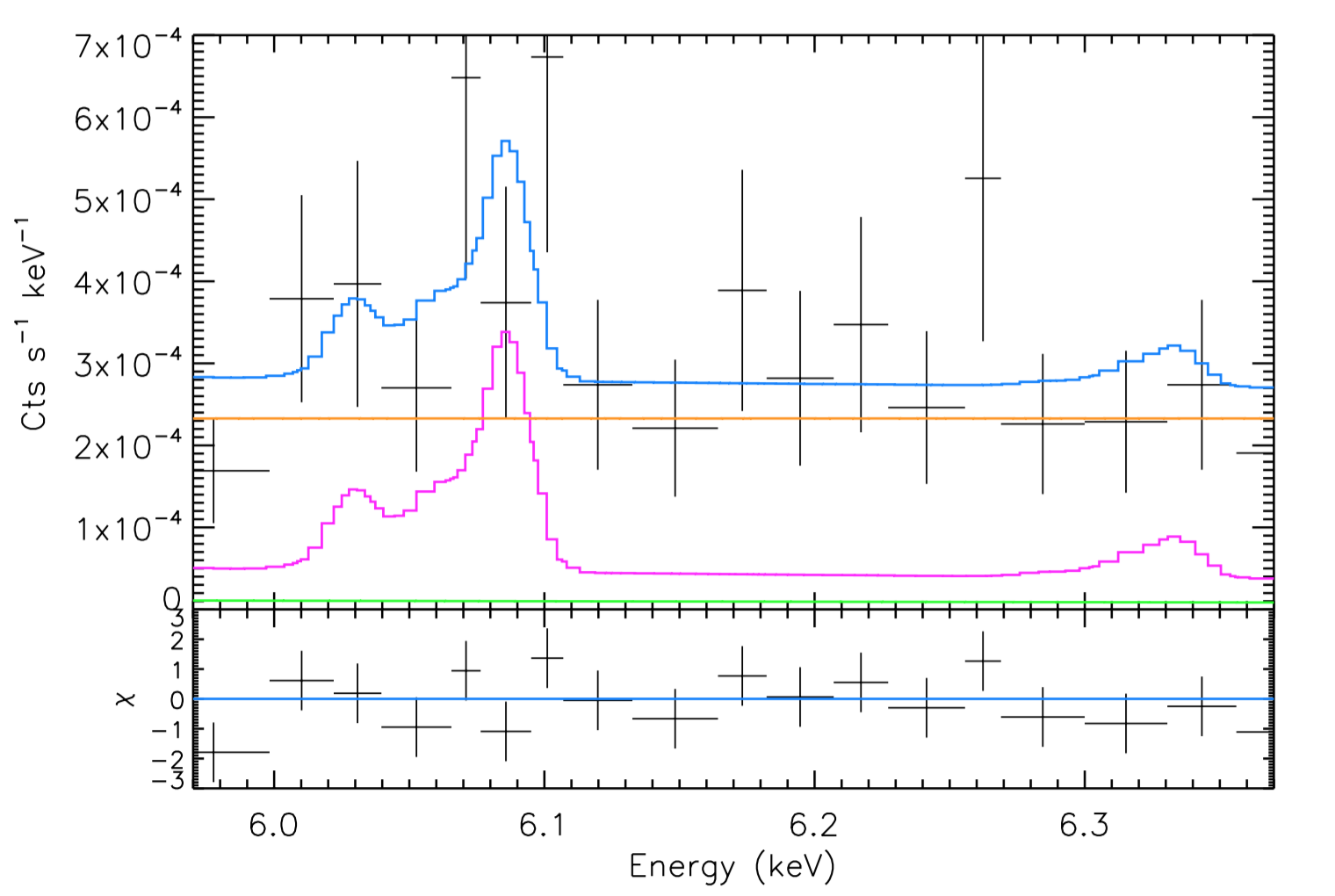}
\caption{Left panel: X-IFU simulated spectrum of the gas bridge and its spectral analysis. Spectral data points with error bars (black crosses) are shown with the best-fit model (blue solid line) and its three components: thermal emission from the gas (magenta), X-ray astrophysical background (green) and particle background (orange). Fit results, with errors, for the four physical quantities of the gas components are also shown. The bottom subpanel shows the residuals with respect to the model. Right panel: same as left panel, but zooming on the 6--6.4 keV energy range (highlighted in yellow in the left panel) where the most prominent emission lines (blended Fe~{\sc xxv} and Fe~{\sc xxvi} K complexes) are present. In both panels data points have been rebinned for display purposes. The plot scaling for the spectra is logarithmic and linear in the left and right panels, respectively.
}
\label{fig_xifu}
\end{figure*}

These first results offer  an interesting physical application to the study of shock waves and particle acceleration in the periphery of galaxy clusters: namely a new method to constrain the shock Mach number based on spectroscopic analysis. 
If we assume that the measured velocity dispersion is similar to the velocity jump induced by the shock, $w \approx \Delta v'=\Delta v/\cos \phi$ (in which $\phi$ is the inclination of the shock normal with respect to the line of sight), and that the local sound speed is given by the temperature probed by the spectroscopy, through the "velocity-jump" method  \citep[][]{va09shocks} we derive

\begin{equation}
\mathcal{M_{\rm XIFU}} = \frac{2}{3} (\frac {\Delta v'}{c_s} + \sqrt{\frac{4\Delta v'}{c_s}+9})
    \end{equation}
which yields $\mathcal{M_{\rm XIFU}} \approx 2.42$ for the X-IFU best fit values of $T \approx  5.33~\rm keV$,  $w \approx 465 ~\rm km/s$ and considering $\phi \approx 75^{\circ}$ for the shock normal (as suggested by Fig.~\ref{fig_rot}, right panel).  This is is not too off from the $\mathcal{M} \sim 2.5-3$ range of Mach number we can measure in 3D within the X-IFU's Field of View, and suggests the interesting possibility of an independent way of constraining the shock parameters via spectroscopic analysis, to be combined and compared with the available radio information there (e.g. Mach number estimated from radio spectral indices for high resolution observations and particle acceleration efficiency).
We comment that intracluster bridges, given their particular geometrical selection, may allow performing this study in a more robust way than in more internal regions of galaxy clusters, where radio relics are found, because in these external regions the local sound speed is low enough, and the velocity field is large enough, that the Doppler broadening due to shocks with a large inclination along the line of sight may be clearly detected via spectroscopic analysis.


\begin{table}
\caption{Physical properties of the plasma computed in the simulation volume compared to the ones measured through the end-to-end X-IFU simulation, with spectral fitting, for a 1Ms integration. The reference (input) temperature is the spectroscopic-like, $v$ and $w$ are the emission weighted averages and $A$ is the metal abundance, computed in the simulation volume as in \protect\cite{roncarelli18}.}
\centering \tabcolsep 5pt 
\begin{tabular}{c|c|c}
Field & Input (sim) & Output (X-IFU)\\  \hline
$T$ [keV] & 5.52 & 5.33 $\pm$ [-0.31, 0.40]  \\
$v$ & 65 & 215 $\pm$ [-146, 123]\\
$w$ &   465 &  384 $\pm$ [-148, 144]\\
$A$ [1/$Z_\odot$] & 0.20 & 0.25 $\pm$ [-0.05, 0.06]\\
\end{tabular}
\label{tab_z}
\end{table}


\section{Discussion: Physical and Numerical uncertainties}
\label{sec:discussion}

Our results are based on non-radiative MHD simulations, in which the role of galaxy formation processes are neglected. This introduces a number of caveats due to the lack of energy losses (e.g. radiative cooling processes) and feedback processes. 

In general, we expect the most relevant effects of cooling and feedback to be limited to cluster centres. While for  $\leq 0.2 ~R_{\rm 100}$ the combined effect of cooling, star formation and feedback (stellar or AGN ones) can introduce a significant amount of clumping in clusters \citep[][]{ro06,nala11},  as well as increase the gas density in cluster cores  $\sim 10$ times compared to non-radiative simulations. However, such effects are predicted to be negligible at distances larger than $\sim R_{\rm 100}$ \citep[e.g.][]{va13feedback}, as well as within intracluster bridges.\\

The absence of galaxy formation processes limits our ability to model the 3D distribution of metals in the outer regions of galaxy clusters. Hence, we assume a uniform $Z=0.3 Z_{\odot}$ gas metallicity. The assumption of uniform metallicity for all elements reduces the number of free parameters making the spectral fitting presented in Section~\ref{subsubsec:sixte} easier. In reality, as pointed out by \cite{cucchetti18}, the X-IFU spectral analysis will require a detailed treatment of the various emitting ions due to the chemical complexity of the ICM \citep[see also][]{roncarelli10,biffi13}. However, most of the information on $w$ is encoded in the Fe~{\sc xxv} and Fe~{\sc xxvi} K complexes lines (see the right panel of Fig.~\ref{fig_xifu}), so this simplification has a minor impact on the accuracy of our results.

\begin{figure*}
  \includegraphics[width=0.495\textwidth]{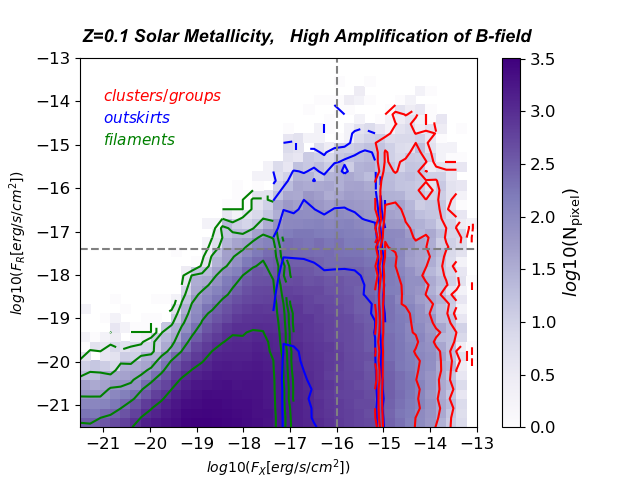}
    \includegraphics[width=0.495\textwidth]{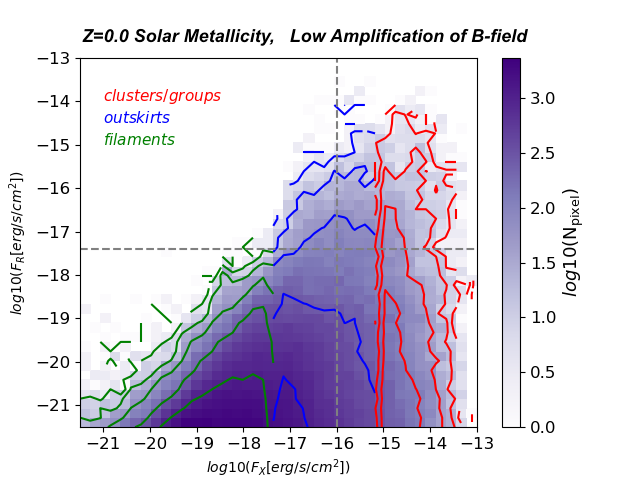}
\caption{Distribution of X-ray and radio flux for variations of our sky model. In the left panel we considered a "high amplification model" for the magnetic field and we assumed a uniform gas metallicity of $Z=0.1 ~Z_{\odot}$ everywhere (instead of $Z=0.3~Z_{\odot}$ as in our baseline model); in the right panel we considered a low-amplification model and a zero metallicity everywhere. The meaning of colors and contours are as in Fig.~\ref{fig:phase}. See text for more details on the model variations. }
\label{fig:phase2}
\end{figure*}

In Fig.~\ref{fig:phase2} we present a few variations of our model, concerning the assumed gas metallicity or the magnetic field strength in our simulation. 

The comparison of the two panels of Fig.~\ref{fig:phase2} with Fig.~\ref{fig:phase} 
shows that the impact of  metal line emission on the detectability of clusters 
outskirts is small, as the blue contours barely change. The reason for this is that in the $\sim 10^7 \rm ~K$ temperature range of intracluster bridges the line emission only accounts for a 
few $\%$ of the total X-ray emission in the $0.8-1.2$ keV band of {\athena}-WFI. 

More critical is the level of gas metallicty for any spectrosopic attempt of characterizing the local plasma condition, as in our mock X-IFU observation described in Sec.~3.3.4. 
However, in this case we used already the more conservative value of $0.2 ~ \rm Z_\odot$ (spatially uniform everywhere) for our 1 Ms mock XIFU observation. Also in this case, a robust measurement of local plasma parameters is possible, provided that the particle and instrumental background are understood. Our tests also indicate that conversely if $Z \geq 0.1 Z_{\odot}$ in intracluster bridges, the reconstruction process of gas conditions through spectroscopic  analysis (even with a 1 Ms integration) will be unreliable and dominated by large uncertainties.
However, a $\sim 0.1 Z_{\odot}$ metallicity is probably even too conservative, and that larger metallicity values should be there \citep[e.g.][]{2018SSRv..214..123B,2018SSRv..214..129M}.

Recently, \citet{2019MNRAS.482.4972K} reconsidered the contribution from the resonantly scattered cosmic X-ray background to the line emission for the  WHIM. 
Resonant scattering can increase the emissivity of the WHIM, considered in this work,
by a factor $\sim 30$. 
However, this boost is limited to the gas at $T \leq 10^6 ~\rm K$ and, 
when integrated in the [0.5-1] keV band, it is of order of $\sim 4$ 
for the coldest part of the WHIM only. 
Therefore, this effect is not expected to contribute to the detectability of the much hotter gas located in intracluster bridges.

\bigskip


Next, we we tested realistic variations of the magnetic field model, which affects the level of predicted synchrotron emission (see Fig.~\ref{fig:phase2}).

Magnetic fields may in principle be overstimated in cosmic filaments, in case the seed field are not of primordial origin and/or there is no dynamo amplification capable to increase the magnetisation of the WHIM to the $\sim 10 ~\rm nG$ level \citep[][]{va17cqg}.  Conversely, limited to the environment of intracluster bridges, our AMR simulation can resolve ongoing dynamo amplification of seed fields, albeit with final fields strengths which are typically far from equipartition with the kinetic energy, at least in our simulation \citep[e.g.][]{lo18}.
The contribution from un-resolved gas motions by the finite numerical resolution in our scheme may underestimate the level of small-scale dynamo amplification, which gets independent from the ampitude of seed fields for large enough Reynolds number \citep[e.g.][]{cho14,2016ApJ...817..127B}. If this is the case, the $\sim 0.1-0.2 \rm ~\mu G$ we measured for our intracluster bridge in Sec.3.3.4 may be underestimated, even if to our knowledge this run is the most resolved so far for objects of this kind. 

To bracket uncertainties, we followed \citet{va15radio} and used a post-processing method to model two extreme scenarios for the amplification of magnetic fields in the cosmic web: a high-amplification model (HA) and a low-amplification model (LA).  In both scenarios, the magnetic field strength from the simulation is renormalized a-posteriori, depending on the local gas overdensity. In the HA model, we account for a the efficient magnetisation of all cosmic gas denser than the critical gas mean density ($\rho \geq \langle \rho \rangle$), which cannot be resolved in the simulation. Conversely, in the LA case we assume that the amplification can be efficient only for densities larger than that of virialized halos ($\rho \geq 300 \langle \rho \rangle$). In both cases, the post-processing normalization is such that the magnetic field energy within each cell becomes $=1\%$ of the cell thermal energy, if the gas density in the cell falls within the HA or LA overdensity range, or remains the original one from our baseline MHD simulation otherwise. As discussed in \citet{va15radio}, the two models produce very different radio fluxes, while they yield similar fluxes in the inner parts of clusters. The two panels in Fig.~\ref{fig:phase2} show that the radio flux from filaments is $\geq 10$ times larger if magnetic fields are much more amplified in filaments as our simulation can resolve. On the other hand, the predicted radio emission from cluster outskirts and from the innermost cluster regions would be only be mildly increased (factors $\leq 4-5$ on average) because for most of radio bright shocks there the magnetic energy  in our simulation is not far from being  $\sim 1\%$ of the thermal gas energy there. 

 For the electron acceleration, our model only includes {\it direct} acceleration of electrons from the thermal pool \citep[][]{hb07}, but it neglects the effect of shock re-accelerated electrons \citep[][]{2013MNRAS.435.1061P,ka12}. However, the density of fossil electrons in cluster outskirts and in bridges is largely uncertain and any level of fossil electrons will increase the emission beyond our estimates.

Also, in our model we neglected for simplicity any further injection or re-acceleration of electrons by additional processes (e.g. turbulent reacceleration, re-connection, shock drift acceleration etc. e.g. \citealt{bj14} and \citealt{by19} for recent reviews) which may power the emission, expecially at low frequency, beyond our estimates.\\

Recent LOFAR-HBA observations suggest indeed that more volume filling and diffuse radio emission processes may be present on $\sim$ $ \rm Mpc$ scales outside of galaxy clusters (Govoni et al., submitted). 

Given the above issues, the X-ray and radio emission model considered in the main paper represents a conservative {\it lower limit} on the joint detectability of intracluster bridges and extreme cluster outskirts.




\section{Conclusions}
\label{sec:conclusions}

We have presented mock radio and X-ray observations of the comic web, based on recent cosmological simulations obtained with the cosmological MHD code {\enzo} \citep[][]{enzo14}. 
Extending our first exploratory study in \citet{athena_ska},  we quantified the  chances of "double detecting" the shocked cosmic web in both frequency ranges, and propose best observing strategies tailored for
future instruments (e.g. SKA and {\athena}). 
Our study highlights that the most promising targets for double detections outside of galaxy clusters are typically located in matter "bridges" connecting pairs of galaxy clusters in an earlier merger state. At this interaction stage, both radio and X-ray emission are boosted compared to the more typical conditions found in cluster outskirts. Such (transient) excess emissions appear to be within the detection range of existing (LOFAR, MWA, ASKAP) and future (SKA-LOW and SKA-MID) radio surveys, as well as of very long ($\geq 100$ ks) integrations with {\athena}, XMM and eROSITA. 
Based on our simulations, the chances of double detections get greatly increased for pairs of objects with a physical (3D) association, with masses in excess of $\geq 10^{14}$  $\rm M_{\odot}$ and with a projected separation between 1 and 4 mean virial diameters (e.g. the sum of the two virial radii of interacting clusters).
For practical purposes, a prior detection of such bridges in the radio domain is expected to serve a strong indication of the possibility of detecting emission also in the soft ($\leq 1.2-2$ keV) X-ray band.\\

Detecting radio emission from transient shocks in such systems will also represent a strong motivation to attempt  long dedicated integrations with  \athena's X-IFU. For example, a deep ($\sim 1 \rm ~Ms$) integration with X-IFU on such jointly detectable portion of intracluster bridges should represent a new strong scientific case to study plasma shock physics in the rarefied environment of the WHIM, namely by  allowing the derivation of the shock Mach number, entirely from spectroscopically-derived information of the local gas velocity dispersion and of the local sound speed, in a temperature regime which is difficult to find in galaxy clusters (Sec.~3.3.4).\\ 

Closely interacting pairs of galaxy clusters have already been detected, and  observations have highlighted unexpected thermal and non-thermal gas features in the interaction region of galaxy clusters in an early merging state \citep[e.g.][]{2017A&A...606A...1A,2017PASJ...69...93S,2017MNRAS.472.2633C,2018MNRAS.478..885B,2018MNRAS.478.2927B,2018ApJ...858...44A}.  Also Sunayaev-Zeldovich observations of close pairs have hinted the presence of dense and X-ray undetectable gas in such associations \citep[][]{2013A&A...550A.134P,2018A&A...609A..49B}. 

Such objects are clearly different from cosmic filaments that are produced by simulations on much larger scales \citep[e.g.][]{2016MNRAS.462..448G}. Moreover, the gas in these bridges is on a thermodynamical state which is different from  the standard WHIM, as the typical density and temperature values are a factor $\sim 10$ larger than in the WHIM, and more into the ICM ballpark. 
However, simulations suggest that bridges are relatively short-lived ($\leq \rm ~Gyr$),  and should undergo a fast evolution compared to filaments on a larger scale, and used to be part of the standard WHIM a $\sim \rm ~Gyr$ ago, before becoming observable in the X-ray band. 
For this reason they have  the potential of illuminating an important intermediate stage in the evolution of cosmic baryons, where gas that has been only shock-heated once in the past gets fast advected onto larger clusters and is subject to large-scale mixing, reprocessing by weaker shocks and supersonic turbulence. This leads to a transient, "boosted" WHIM phase, with a mean temperature beyond the canonical (but not entirely physically motivated) temperature bounds associated to the WHIM ($10^{5} \rm K \leq T \leq 10^{7} \rm K$). \\

\begin{figure}
   \includegraphics[width=0.49\textwidth]{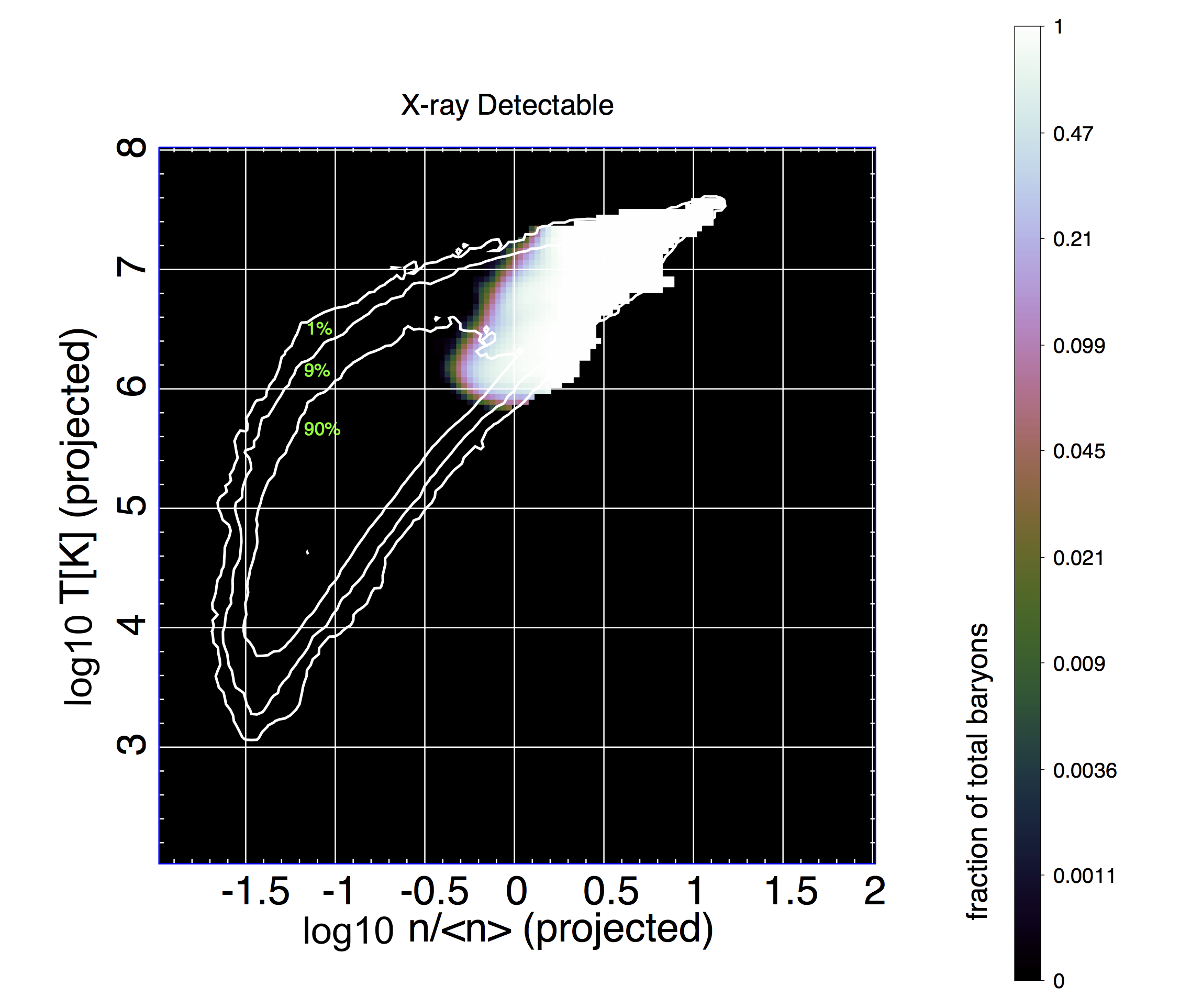}
    \includegraphics[width=0.49\textwidth]{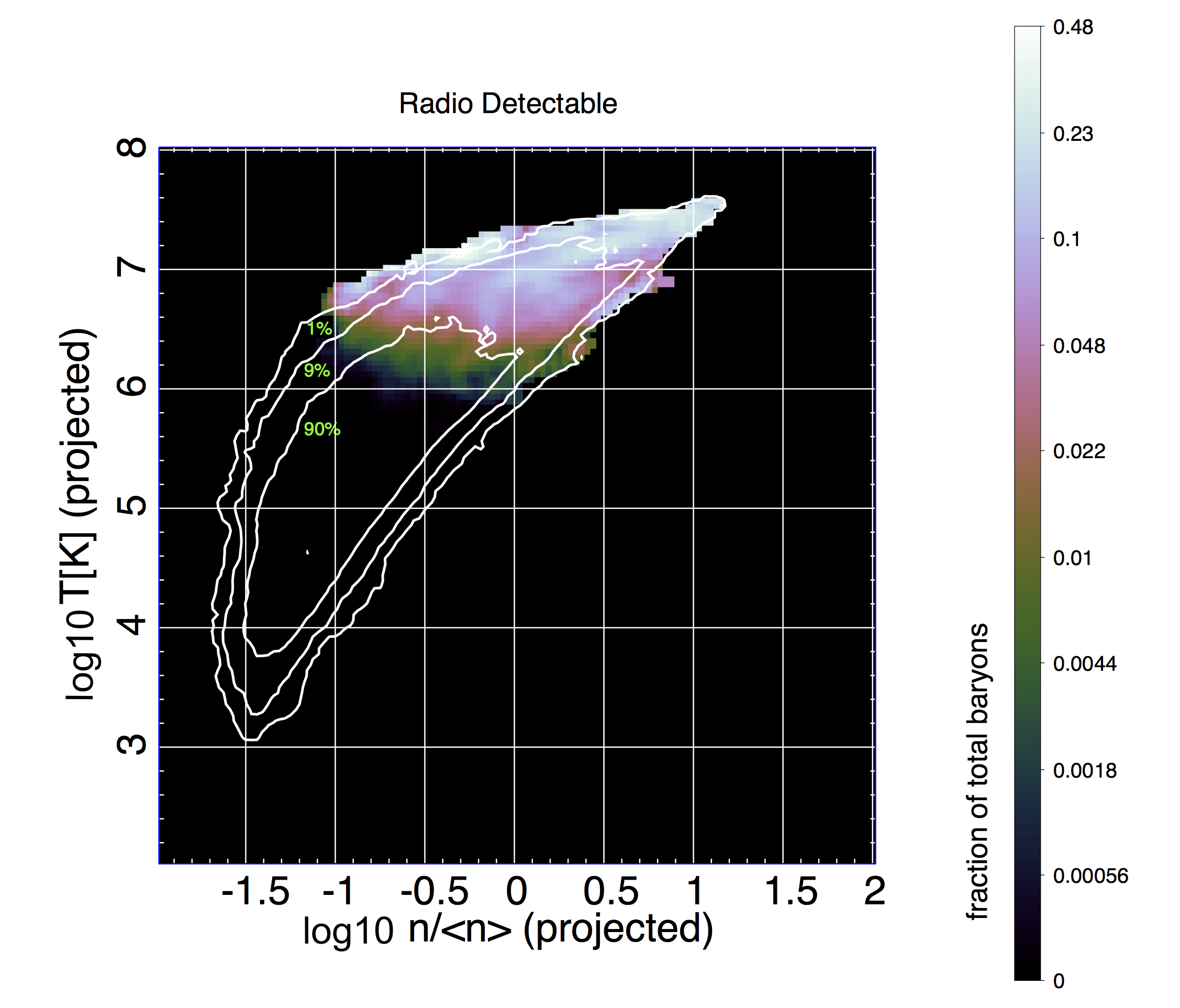}
  \caption{Distribution of projected mean temperature and density for all pixels  in our sky model at $z=0.05$. The total area within the isocontours mark the area where the $90\%$, $9\%$ and $1\%$ of the baryon budget is contained. The colors marks the fraction of the total baryon budget which can be detected with X-ray detections with \athena\ (top panel) or with radio detections with SKA-LOW (bottom panel).}
  \label{whim}
\end{figure}

In summary, our work can stress the importance of the radio band to study the missing baryons of the cosmic web. We quantify this by presenting in Fig.~\ref{whim} the distribution of the mean temperature and gas over density  for all pixels  in our sky model  {\footnote {As noted in 3.1, the mass-weighted temperature and volume-weighted gas density here 
{\it underestimate} the corresponding three-dimensional values (where most of the emission along the LOS is produced) by $\sim 10$.}}. Contours denote the total fraction of baryons at $z=0.05$. Overlaid is the fraction of baryons that should be detectable by X-ray observations using a 1 Ms exposure with \athena,  or with the SKA-LOW survey at $260$ MHz. \\

Clearly, X-ray observations are most efficient in the high-temperature and high-density part of the plot. However, less than $10 \%$ of the total budget of baryons in the Universe are located in this range. Conversely, radio surveys can  trace only a much a smaller fraction of hot and dense baryons in the Universe, due to small filling factor of shocks leading to radio emission in this regime. 
However, radio detections trace baryons with significantly lower projected temperature and density compared to X-ray observations, enabling them to probe into the gas phase where  $\sim 90\%$ of cosmic baryons are. As such detections can only illuminates the {\it shocked} portion of the WHIM (or immediately downstream of it), it will be crucial to assess the bias factor between the radio emitting fraction of the cosmic web, and its larger ("radio quiet") component. 
With the assistance of advanced numerical simulations, capable of assessing this bias as a function of environment, wide and deep radio surveys  will have the potential to convert systematic detections of radio shocks in the rarefied cosmic web into an estimate of the amount of missing baryons in Universe. 
 
\begin{acknowledgements}

We thank our anonymous referee for the fruitful feedback on the first version of the paper, which has led to an improved presentation of our results. 
The cosmological simulations were performed with the {\enzo} code (http://enzo-project.org), which is the product of a collaborative effort of scientists at many universities and national laboratories. We gratefully acknowledge the {\enzo} development group for providing extremely helpful and well-maintained on-line documentation and tutorials.
F.V. acknowledges financial support from the ERC  Starting Grant "MAGCOW", no. 714196.   
We acknowledge the  usage of computational resources on the Piz Daint supercomputer at CSCS-ETHZ (Lugano, Switzerland) under projects s701 and s805 and at the J\"ulich Supercomputing Centre (JFZ) under project HHH42. We also acknowledge the usage of online storage tools kindly provided by the Inaf Astronomica Archive (IA2) initiave (http://www.ia2.inaf.it). 
S. E. and M. R. acknowledge funding from the European Union's Horizon 2020 Programme under the AHEAD project (grant agreement n. 654215).
S.E. acknowledges financial contribution from the contracts NARO15 ASI-INAF I/037/12/0, ASI 2015-046-R.0, and ASI-INAF n.2017-14-H.0.
We acknowledge fruitful scientific feedback by A. Bonafede,   M. Cappi, M. Markevitch, N. Locatelli, R. Cassano, G. Brunetti, I. Prandoni, E. Churazov and I. Khabibullin, which improved the presentation of our results.  We also thank T. Boller and V. Ghirardini to have provided useful information on the performance of eROSITA and XMM-Newton, respectively. 

\end{acknowledgements}

\bibliographystyle{aa}
\bibliography{franco}

\end{document}